\documentclass[a4paper,fleqn,usenatbib]{mnras}
\pdfoutput=1

\usepackage[T1]{fontenc}
\usepackage{ae,aecompl}
\usepackage{epstopdf}
\usepackage{aas_macros}
\usepackage{braket}
\usepackage{verbatim}
\usepackage{wrapfig}
\usepackage{booktabs}
\usepackage{amsfonts}
\usepackage{amsmath}
\usepackage{appendix}
\usepackage{graphicx}
\usepackage[usenames,dvipsnames]{color}
\usepackage[normalem]{ulem}
\usepackage{array}
\newcommand{\hMpc}{h^{-1} {\rm Mpc}} 
\newcommand{\camb}{\textsc{camb}}

\newcommand{\halofit}{\textsc{halofit}}

\title[WiggleZ 2D BAO]{Measuring the 2D Baryon Acoustic Oscillation signal of galaxies in WiggleZ: Cosmological constraints} 

\author[S. R. Hinton et al.]{Samuel R. Hinton,$^{1,2}$\thanks{E-mail: \href{samuelreay@gmail.com}}
Eyal Kazin,$^{3,2}$
Tamara M. Davis,$^{1,2}$
Chris Blake,$^{3,2}$\newauthor
Sarah Brough,$^{5}$
Matthew Colless,$^{4}$
Warrick J. Couch,$^{5}$ \newauthor
Michael J. Drinkwater,$^{1}$
Karl Glazebrook,$^{3}$
Russell J. Jurek,$^{6}$
David Parkinson,$^{1,2}$ \newauthor
Kevin A. Pimbblet,$^{7,8,9}$
Gregory B. Poole,$^{10}$
Michael Pracy$^{11}$ and
David Woods$^{12}$
\\
$^{1}$School of Mathematics and Physics, The University of Queensland, Brisbane, QLD 4072, Australia\\
$^{2}$ARC Centre of Excellence for All-sky Astrophysics (CAASTRO)\\
$^{3}$Centre for Astrophysics \& Supercomputing, Swinburne University of Technology, P.O. Box 218, Hawthorn, VIC 3122, Australia\\
$^{4}$Research School of Astronomy and Astrophysics, The Australian National University, Canberra, ACT 2611, Australia\\
$^{5}$Australian Astronomical Observatory, PO Box 915, North Ryde, NSW 1670, Australia\\
$^{6}$Australia Telescope National Facility, CSIRO, Epping, NSW 1710, Australia\\
$^{7}$E.A.Milne Centre for Astrophysics, University of Hull, Cottingham Road, Hull HU6 7RX, UK\\
$^{8}$School of Physics, Monash University, Clayton, Victoria 3800, Australia;\\ 
$^{9}$Monash Centre for Astrophysics (MoCA), Monash University, Clayton, Victoria 3800, Australia\\
$^{10}$School of Physics, University of Melbourne, Parksville, VIC 3010\\
$^{11}$Sydney Institute for Astronomy, School of Physics, University of Sydney, NSW 2006, Australia\\
$^{12}$Department of Physics and Astronomy, University of British Columbia, 6224 Agricultural Road, Vancouver, BC V6T 1Z1, Canada
}
\date{Accepted XXX. Received YYY; in original form ZZZ}

\pubyear{2016}

\begin{document}

\label{firstpage}
\pagerange{\pageref{firstpage}--\pageref{lastpage}}
\maketitle

\begin{abstract}
We present results from the 2D anisotropic Baryon Acoustic Oscillation (BAO) signal present in the final dataset from the WiggleZ Dark Energy Survey.  We analyse the WiggleZ data in two ways: firstly using the full shape of the 2D correlation function and secondly focussing only on the position of the BAO peak in the reconstructed data set. When fitting for the full shape of the 2D correlation function we use a multipole expansion to compare with theory.  When we use the reconstructed data we marginalise over the shape and just measure the position of the BAO peak, analysing the data in wedges separating the signal along the line of sight from that parallel to the line of sight. 

We verify our method with mock data and find the results to be free of bias or systematic offsets.  We also redo the pre-reconstruction angle averaged (1D) WiggleZ BAO analysis with an improved covariance and present an updated result.   The final results are presented in the form of $\Omega_c h^2$, $H(z)$, and $D_A(z)$ for three redshift bins with effective redshifts $z = 0.44, 0.60$, and $0.73$. Within these bins and methodologies, we recover constraints between 5\% and 22\% error. Our cosmological constraints are consistent with Flat $\Lambda$CDM cosmology and agree with results from the Baryon Oscillation Spectroscopic Survey (BOSS).
\end{abstract}

\clearpage
\section{Introduction}

Modern cosmological observations have given strict constraints on cosmological parameters and model viability, and indicate a late time accelerated expansion of the universe \citep{RiessFilippenko1998, PerlmutterAldering1999, SpergelVerde2003, RiessStrolger2004, TegmarkBlanton2004, SanchezBaugh2006, SpergelBean2007, Komatsu2009, RiessMacri2009, PercivalReid2010, ReidPercival2010,BlakeKazin2011}. Determining the cause of this accelerating expansion is one of the foremost problems in cosmology.   Continued efforts to measure the expansion history of the universe and growth of structure within it will allow differentiation between many proposed models such as those that invoke ``dark energy'' and those that invoke a modification to general relativity \citep{SanchezScoccola2012}. One area of rapid development is using Baryon Acoustic Oscillations (BAO) measured in the large scale structure of the universe to provide a robust and precise measurement of the history of the universe's expansion rate and size \citep{EisensteinHu1998,BlakeGlazebrook2003,HuHaiman2003,Linder2003,SeoEisenstein2003}. Analysis of the BAO signal has been performed on several cosmology surveys, providing tight constraints on cosmological parameters \citep{Eisenstein2005,PercivalCole2007,Gaztanaga2009,PercivalReid2010,BlakeDavis2011,BlakeKazin2011, SanchezKazinBeutler2013, AndersonAubourg2014}. The constraints BAO measurements provide are highly complementary to, and can be used in conjunction with, constraints derived from measurements on the Cosmic Microwave Background  \citep[CMB;][]{BennettHalpern2003, Planck201416}, weak lensing \citep{VanWaerbeke2000,WittmanTyson2000,KaiserWilson2000}, and supernova data \citep{KowalskiRubin2008, KesslerBeckerCinabro2009, BetouleKessler2014}.

Here we assess the 2D galaxy correlation function, which groups pairs of galaxies by their angle with respect to the line of sight.  The correlation function of galaxy pair separations along the line of sight is most sensitive to the Hubble parameter, $H(z)$, and  perpendicular to the line of sight is more sensitive to the angular diameter distance, $D_A(z)$.  

Decomposing the BAO signal into the line of sight and tangential components has only recently become possible \citep{Gaztanaga2009, XuCuesta2013, AndersonAubourg2014DR11, AndersonAubourg2014}. In addition to fitting for the 2D BAO signal in the full shape of the galaxy correlation function (including BAO), reconstruction techniques have recently been utilised to recreate a stronger BAO peak at the expense of marginalising over the broad shape \citep{PadmanabhanXuEisenstein2012, KazinKoda2014}.

In this paper we analyse the 2D BAO signal using both techniques on the WiggleZ Dark Energy Survey data.  In detail we:
\begin{enumerate}
\item {\bf Model the multipole correlation function:} We use the full shape information in the correlation function by modelling its multipoles and fitting it to multipole data extracted from the WiggleZ survey.  This uses the maximal information in the correlation function, but does not include reconstruction and therefore has a weaker BAO peak. 
\item {\bf Use reconstruction and only measure the BAO peak:}  We perform reconstruction on the WiggleZ data, which recovers a correlation function with a much stronger BAO peak, but loses the shape information.  We therefore marginalise over the shape information and only use the peak itself as a standard ruler. 
\end{enumerate}

In this paper Section \ref{sect:wigglez} begins by describing the WiggleZ data and WizCOLA simulations we utilise, and details relevant previous studies that make use of the datasets.  Then in Section \ref{sec:model} we construct a theoretical model of the full 2D correlation function and we decompose that correlation function into our two summary statistics --- multipole expansion and wedges.
Section \ref{sec:test} evaluates those models against the WizCOLA simulations, and Sections \ref{sec:multi} and \ref{sec:wedge} use the multipole and wedge models to extract cosmological parameters from the unreconstructed and reconstructed WiggleZ data respectively. In Section \ref{sec:disc} we place these results into the larger cosmological context by incorporating the results from other surveys \& other methodologies, and present final conclusions.

\begin{figure*}
	\begin{center}
		\includegraphics[width=\textwidth]{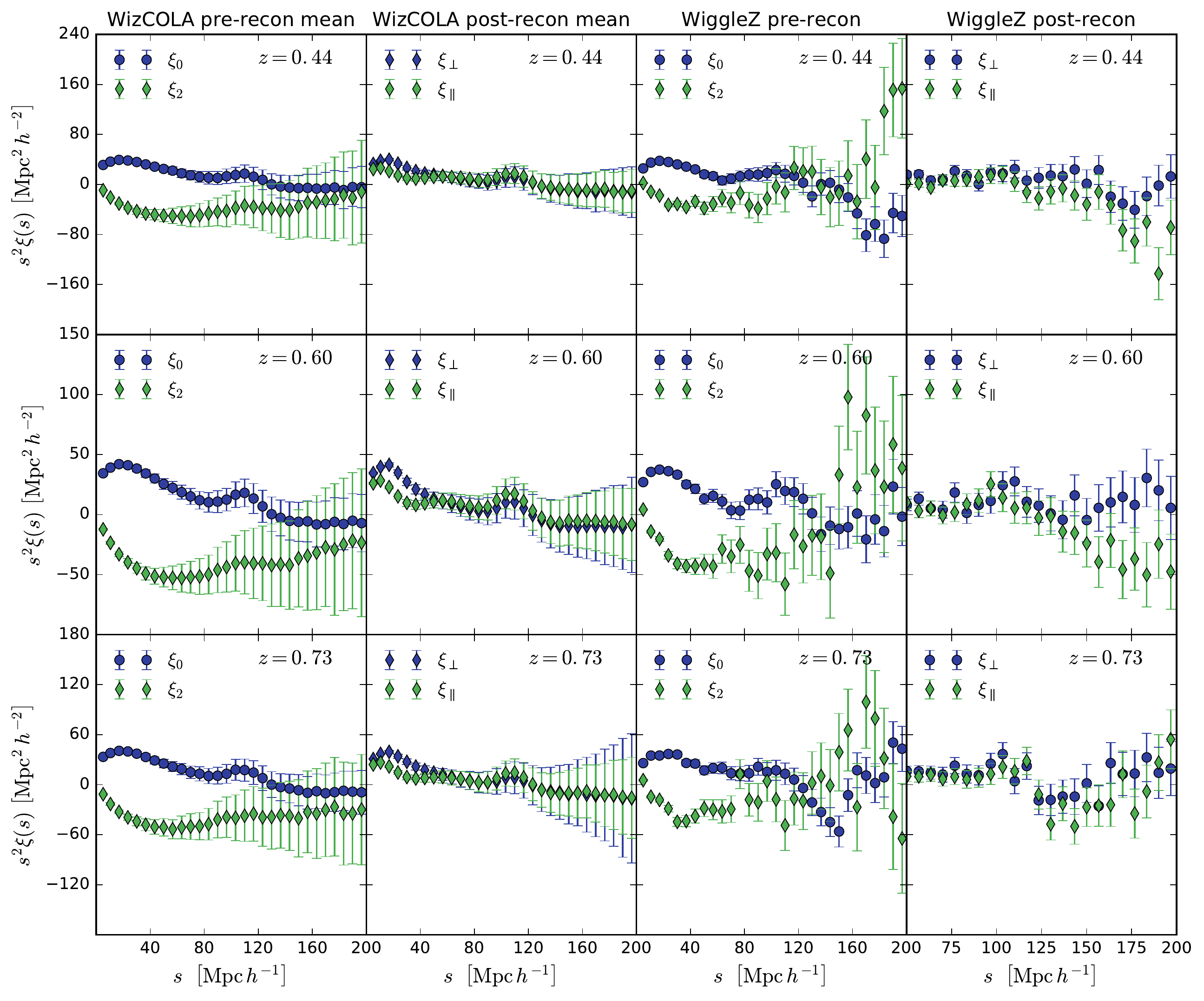}
	\end{center}
	\caption{The WizCOLA simulation mean data \citep{KazinKoda2014,KodaBlake2015} (both pre-reconstruction multipoles and post-reconstruction wedges), unreconstructed WiggleZ multipoles, and reconstructed WiggleZ wedges shown in four respective columns. Rows represent different redshift bins in the data. For the multipole data, the monopole contribution and quadrupole contribution are shown in blue circles and green diamonds respectively. For the wedge data, we show the transverse component (blue circles) and line-of-sight component (green diamonds). Uncertainty is determined by looking at simulation variance over 600 realisations of fiducial cosmology. For details on multipole and wedge constructions of the 2D correlation function, see \S\ref{sec:model}.}
	\label{fig:wizcola}
\end{figure*}

\section{The WiggleZ Dark Energy Survey}\label{sect:wigglez}

The WiggleZ Dark Energy Survey was carried out between 2006 to 2011 at the Australian Astronomical Observatory over the course of 276 nights \citep{Drinkwater2010}. The survey measured redshifts of $225\,415$ galaxy spectra, targeting blue emission-line galaxies in a redshift range of $0.2 < z < 1.0$. The target selection function is summarised in \citet{BlakeDavis2011}, and explained in detail in \citet{BlakeBrough2010}.

A variety of analyses have already been conducted on the WiggleZ dataset. 
The one dimensional BAO signal was analysed for a subset of WiggleZ data in \citet{BlakeDavis2011}, and this analysis was refined by both including the full survey data and subdividing the data into redshift bins in \citet{BlakeKazin2011}. A final analysis of the 1D BAO signal involving reconstruction of the BAO peak was performed by \citet{KazinKoda2014}. 

Analyses of the 2D data have also been performed on WiggleZ data, but not yet on scales large enough to include the BAO peak.  \citet{BlakeBroughColless2011} and \citet{ContrerasBlake2013} use redshift space distortions to measure the rate of growth  of structure, while \citet{BlakeGlazebrook2011} used the Alcock-Paczynski test on galaxy clustering as a {\em standard sphere} to measure expansion history. Cosmological results from the WiggleZ papers were combined with other surveys and datasets in \citet{Parkinson2012} and analysis of the overlap regions wit the Baryon Oscillation Spectroscopic Survey (BOSS) were completed in \citet{BeutlerBlakeKoda2016} and \citet{MarinBeutlerBlake2016}.

One investigation that has not been undertaken with the WiggleZ data is a two dimensional analysis of the BAO signal.  As the survey meets the criteria for being able to detect the 2D BAO signal -- volumes of order 1 Gpc$^3$ with order of $10^5$ redshifted galaxies \citep{Tegmark1997,BlakeGlazebrook2003,BlakeParkinson2006} -- we present that analysis in this paper. The high-redshift range and low-bias galaxy selection of the WiggleZ survey makes such an analysis a useful consistency check on larger current surveys such as BOSS \citep{AndersonAubourg2014}.

This analysis is motivated by two recent improvements to the WiggleZ survey data.  The first improvement is that reconstruction has now been performed to remove some of the effect of peculiar velocities \citep{KazinKoda2014}, which sharpens the BAO peak and thus makes it  easier to measure in the 2D correlation function (previous analyses had amplified the signal by averaging the information across all angles).  The second improvement is the creation of accurate mock catalogues from the WizCOLA simulations \citep{KodaBlake2015} for both the pre- and post-reconstruction data. The simulations provide more accurate covariance estimates than the lognormal realisations used in the early analyses.  We therefore also revisit the pre-reconstruction angle-averaged (1D) constraints from the final WiggleZ survey and present updated results.  By fitting our theoretical models to the mock data and recovering the correct cosmological model (the model that was used to make the simulations) we are able to perform rigorous checks that our correlation function models are sufficiently accurate, and optimise the range of scales over which the theory is adequate to include in the fits. Figure~\ref{fig:wizcola} shows these improvements - detailing the galaxy correlation function for the WizCOLA mean multipole data, pre-reconstruction WiggleZ multipole data, and the post-reconstruction WiggleZ wedge data.

\section{The 2D Correlation Function}
\label{sec:model}

\subsection{Base Model --- before reconstruction}
For fits to the unreconstructed data, we fit against not just the BAO peak, but also to the broad shape of the correlation function. We begin the model with a linear power spectrum $P_{\rm{lin}}(k)$, which is  generated using the \camb{} software created by \citet{Lewis2000}. We limit our analysis to a Flat $\Lambda$CDM cosmology, appropriate as the data are of insufficient strength to tightly constrain more parameters.
We set $\Omega_c h^2$ as a free parameter, and fix other values to the WizCOLA simulation fiducial cosmology in our analysis, such that $\sigma_8 = 0.812$, $n_s=0.961$, $h = 0.705$. We fix $\Omega_b h^2$ to $0.0226$, as $\Omega_b h^2$ is well constrained by CMB data and variations even up to $5\sigma$ are negligible to the BAO model when testing Flat $\Lambda$CDM cosmology. This value is consistent with the WizCOLA simulation value $\Omega_b h^2 = 0.02266$. 

We model the BAO peak smoothing caused by displacement of matter due to bulk flows with a smoothing parameter \citep{CrocceScoccimarro2008, SanchezBaughAngulo2008, Sanchez2009, BlakeDavis2011, BeutlerBlake2011}. This smoothing parameter takes the form of a Gaussian dampening term which reduces the amplitude of the BAO signal as a function of $k$:
\begin{align} \label{{eq:blake1}}
	P_{\rm{dw}}(k) = e^{-k^2 \sigma_v^2} P_{\rm{lin}}(k) + (1 - e^{-k^2 \sigma_v^2}) P_{\rm{nw}}(k),
\end{align}
where $P_{\rm{nw}}(k)$ is a power spectrum without the BAO signal. Whilst advances in renormalization perturbation theory  \citep[RPT;][]{CrocceScoccimarro2008} allow a theoretical determination of $\sigma_v$ as
\begin{align} \label{eq:sigmav}
\sigma_v^2 = \frac{1}{6\pi^2} \int P_{\rm{lin}}(k)\, dk,
\end{align}
however $\sigma_v$ is set to a free parameter due to inaccuracies in the theoretical determination from non-linear effects. It is important to note that past studies have also used an analogous term to $\sigma_v$, defining $k_* = 1/(\sqrt{2}\sigma_v)$.

In most studies the power spectrum without the BAO signal present is generated using the \verb;tffit; algorithm given by \citet{EisensteinHu1998}. \citet{ReidPercival2010} investigated an alternate method of generating a no-wiggle power spectrum from the linear \camb{} power spectrum in which an 8 node b-spline was fitted to the linear power spectrum, concluding the likelihood surfaces generated when fitting using splines and the algorithm from \citet{EisensteinHu1998} agree well. For our work we introduce a new method to attain a no-wiggle power spectrum $P_{\mathrm{nw}}(k)$ utilising polynomial subtraction. For a comparison of this methodology against the \verb;tffit; algorithm supplied by \citet{EisensteinHu1998} or spline fitting, please see Appendix \ref{app:dewiggle}.

The non-linear effects of gravitational growth are incorporated by using \halofit{} from \citet{Smith2003}, which generates a power ratio $r_{\rm{halo}}$ as a function of $k$ 
\begin{align}
	P_{\text{nl}} = P_{\text{dw}} r_{\text{halo}}.
\end{align}
We take into account galaxy bias $b$ and also follow \citet{BlakeDavis2011} who incorporate extra scale dependent bias derived from the GiggleZ simulations, $B(s)$, into to the model, via
\begin{align}
	\xi_{\rm{galaxy}}(s) = B(s) b^2 \xi(s),
\end{align}
where $B(s) = 1 + (s/s_0)^\gamma$, with $s_0 = 0.32 h^{-1}\,$Mpc and $\gamma = -1.36$.

The Kaiser effect from coherent infall can be modelled simply in Fourier space \citep{Kaiser1987}:
\begin{align} \label{eq:gaztanga1}
	P_{\rm{nl}}(k, \mu) = (1 + \beta \mu^2)^2 P_{g}(k),
\end{align}
where $P_{g}$ is the power spectrum of galaxy density fluctuations $\delta_g$, $\mu$ is the cosine of the angle to line of sight, and $\beta = f/b$ and $f$ is the growth rate of growing modes in linear theory. When reconstructing the BAO signal \citep[see][for details]{PadmanabhanXuEisenstein2012,KazinKoda2014}, the Kaiser effect is corrected for and thus does not have to be inserted into the cosmological model.

Peculiar velocity does not have to be coherent to affect observational cosmology, and the random peculiar velocities of galaxies in clusters, which are related to the cluster mass via the virial theorem, create artifacts known as Fingers of God. In the investigation of growth rate with WiggleZ data, \citet{BlakeBroughColless2011} adopts a Lorentzian model of velocity dispersion with
\begin{align}
	P_{\rm{gal}} = \frac{1}{1 + (k \sigma_V \mu)^2}  P_{\rm{nl}}(k, \mu), \label{eq:lorentzian}
\end{align}
where $\sigma_V$ is the pairwise peculiar velocity dispersion and not to be confused with the $\sigma_v$ term accounting for BAO peak damping. We adopt this in our analysis. For a more complete treatment of the underlying model, see \citet{HintonThesis2015}.

\subsubsection{Moving to a correlation function} \label{sec:prior:cor}

The power spectrum and correlation functions are related to each other via Fourier transform. One dimensional BAO analyses generally look at the angle-averaged correlation function, which is simply the monopole moment. A power spectrum can be decomposed into its multipole components via 
\begin{align}
	P_{\ell}(k) = \frac{2\ell + 1}{2} \int_{-1}^1 P_{\rm{gal}}(k, \mu) \ \mathcal{L}_\ell \  d\mu
\end{align}
where $\mathcal{L}_\ell$ represents the $\ell$'th Legendre polynomial. These multipole components can be turned into correlation functions by Fourier transforming them, giving
\begin{align}
	\xi_\ell(s) = \frac{1}{(2\pi)^3} \int 4\pi k^2 \ P_\ell(k) \ j_\ell(ks)
\end{align}
where $j_\ell(ks)$ are spherical Bessel functions of the first kind. The increased power of small scale oscillations from the non-linear corrections decreases convergence of this function, so we multiply the integrand by a Gaussian factor $\exp(-k^2 a^2)$ to improve convergence, where we found $a=0.5 h^{-1} \rm{Mpc}$ to be the optimal factor to improve computational speed while maintaining accuracy \citep{HintonThesis2015}.  \citep[The results are not sensitive to the exact choice of $a$;][set $a= 1\, h^{-1}\rm{Mpc}$.]{AndersonAubourg2012}

\subsubsection{Multipoles and Wedges}

It is impractical to fit the data to a full 2D correlation function, as the calculation of the covariance matrix is infeasible.  Instead one typically reduces the 2D information into a simplified measure that encapsulates the essential anisotropy.  Two methods by which this can be done are wedges and multipoles.  

The wedges method splits the 2D correlation function into wedges based on angle, and averages the correlation function in that wedge.  One could in principle have many wedges, but for our data (and all previous data) the signal to noise limits us to two wedges -- one taking the half of the data along the line of sight ($\mu\ge0.5$), the other perpendicular to it ($\mu<0.5$), where $\mu$ is the cosine of the angle with respect to the line of sight. The multipole method decomposes the correlation function into multipoles -- with the vast majority of signal being found in the monopole and quadrupole moments.  For extended treatment of the mathematics, see \citet{KazinSanchezBlanton2012, KazinSanchezCuesta2013, SanchezKazinBeutler2013, XuCuesta2013}. 

In all cases we have used a fiducial cosmology to convert observed right ascension, declination, and redshift into distances (separations) and thus generate the correlation function.  So the variables we fit for are not the distances (separations) themselves, but rather the ratio of the distance in the true model to the distance in the fiducial model.  This is achieved by scaling the model distances to give $s_{\rm test}= \alpha s_{\rm model}$.   

Thus the primary variable we fit for is $\alpha$.  Depending on which type of analysis we are performing, $\alpha$ relates to  distances in different ways.  \citet{BlakeDavis2011} show, for example, the degeneracy lines between $\alpha$ and $\Omega_{\rm m}$ and how they change depending on whether you fit to the correlation function shape or power spectrum, or only the BAO peak.  When fitting to the BAO peak, the degeneracy direction lies along a line of constant $r_s/D_V$ (in the 1D case)  where 
\begin{align}
D_V=\left[ (1+z)^2 D_A(z)^2 \frac{cz}{H(z)}\right]^{1/3}
\end{align}
and $r_s$ is the sound horizon at drag epoch, given by
\begin{align}
r_s = \frac{c}{\sqrt{3}} \int_0^{1/(1+z)} \frac{da}{a^2 H(a) \sqrt{1 + (3\Omega_b/4\Omega_\gamma) a}}, \label{eq:rs}
\end{align}
with $\Omega_\gamma = 2.469\times10^{-5} h^{-2}$ for $T_{{\rm cmb}}=2.725\,{\rm K}$ and $\Omega_r = \Omega_\gamma(1 + 0.2271 N_{{\rm eff}})$, where we utilise $N_{\rm eff} = 3.04$.
However, when fitting for the correlation function shape the degeneracy direction lies along a line of constant $A=D_V(z) \sqrt{\Omega_m H_0^2} / zc$, which was a parameter introduced by \citet{Eisenstein2005} for exactly that reason.  Note that $A(z)$ does not depend on $r_s$.

When we update the pre-reconstructed angle averaged 1D measurement of \citet{BlakeKazin2011} we fit for 
\begin{align}
\alpha = \frac{D_V(z)}{D_V^{\prime}(z) },\label{eq:DvDv}
\end{align}
with the prime denoting the value from fiducial cosmology.

For the multipole expansion used in the 2D pre-construction fits, we fit a scaling factor $\alpha$ and warping parameter $\epsilon$ such that
\begin{align}
\alpha(1 + \epsilon)^2 \approx \alpha(1 + 2\epsilon) &\approx \frac{H^\prime(z)}{H(z)} \label{eq:alpha1}\\
\alpha(1 + \epsilon)^{-1} \approx \alpha(1 - \epsilon) &\approx \frac{D_A(z)}{D_A^\prime(z) }. \label{eq:alpha2}
\end{align}

In summary, our model generates a linear power spectrum, with input parameter $\Omega_c h^2$. The damping term $k_*$ (equivalently $\sigma_v$), galaxy bias $b^2$, growth rate $\beta$, and Lorentzian factor $\sigma_V$ are marginalised over, leaving final constraints in the form of $\Omega_c h^2$, $\alpha$ and $\epsilon$. 

\subsection{Base model --- after reconstruction}

\citet{EisensteinSeoSirko2007} proposed that the ``blurring'' of the baryon
acoustic peak due to the large-scale coherent motion of galaxies could
be partially remedied by a procedure of ``linear reconstruction,'' in
which the displacement field $\vec{\psi}$ is estimated from the
observed density field using linear theory, and used to retract
galaxies by $-\vec{\psi}$ to an approximation of their initial
position.  In \citet{KazinKoda2014} we applied density-field
reconstruction to the WiggleZ Survey data, and demonstrated that it
resulted in a sharpening of the acoustic peak in the angle-averaged
correlation function and thereby in improved distance constraints,
with consistent behaviour found in mock catalogues.  We overcame edge
effects and holes within the survey by applying a Weiner-filtering
procedure similar to that presented in \citet{PadmanabhanXuEisenstein2012}.  For
full details of the procedure please refer to \S 2.3 in
\cite{KazinKoda2014}.

We now examine the anisotropic baryon acoustic peak signature present
in the reconstructed WiggleZ density field, marginalizing over the
broadband shape information.  A full description of our procedure is
given in our previous analysis of the SDSS DR9 CMASS galaxies
\citet[][see \S 5.3]{KazinSanchezCuesta2013}.  In brief, we measured the
correlation function of the reconstructed data in two ``clustering
wedges''. 

We fitted the data assuming a BAO template including
quasi-linear corrections based on the renormalized perturbation theory
of \citet{CrocceScoccimarro2008}. This template is distorted in the tangential
and radial directions by parameters $\alpha_\perp$ and
$\alpha_\parallel$ which are given by
\begin{align}
\alpha_\perp &\approx \frac{D_A(z) r^\prime_s }{D_A^{\prime}(z) r_s}, \label{eq:alphaperp}\\
\alpha_\parallel &\approx \frac{H^{\prime}(z) r^\prime_s}{H(z) r_s},\label{eq:alphaparallel}
\end{align}
where the $r_s$ term is present (unlike in the pre-reconstruction $\alpha$) due to the degeneracy direction of fitting only for the BAO peak.

We assume a flat prior in $(\alpha_\perp,\alpha_\parallel)$ between
0.5 and 1.5.  For each clustering wedge we also marginalized over an
amplitude parameter and the coefficients of three additive polynomial
terms, producing a 10-parameter model.  We explore the parameter space
using MCMC chains, and present results for
$(\alpha_\perp,\alpha_\parallel)$, marginalizing over the other 8
parameters.

\section{Validation of unreconstructed multipole analysis}
\label{sec:test}

To validate our model we employ several tests.  Firstly we compare it to past analyses (the 1D WiggleZ results) and then in more detail to simulated data (WizCOLA).  Following that we test two methods by which to combine the information in the different redshift bins.

\subsection{Validation against prior WiggleZ analyses}

We use our model to repeat the 1D BAO analysis using the WiggleZ unreconstructed dataset over the same data range utilised by \citet{BlakeDavis2011} and \citet{BlakeKazin2011}: $10<s<180 h^{-1}\rm{Mpc}$. Our model is very similar to the one used by \citet{BlakeKazin2011}, but differs from theirs by implementation (MCMC methods in comparison to a grid search), dewiggling methodology, covariance matrix, and choice of statistical measures reported \citep[we use maximum likelihood statistics, as opposed to mean statistics used in][]{BlakeDavis2011}. The most important difference in the analysis is that we use the improved knowledge of covariances from the WizCOLA simulations \citep[as compared to the lognormal realisations used in][]{BlakeKazin2011}.

\begin{figure}
	\begin{center}
		\includegraphics[width=\columnwidth]{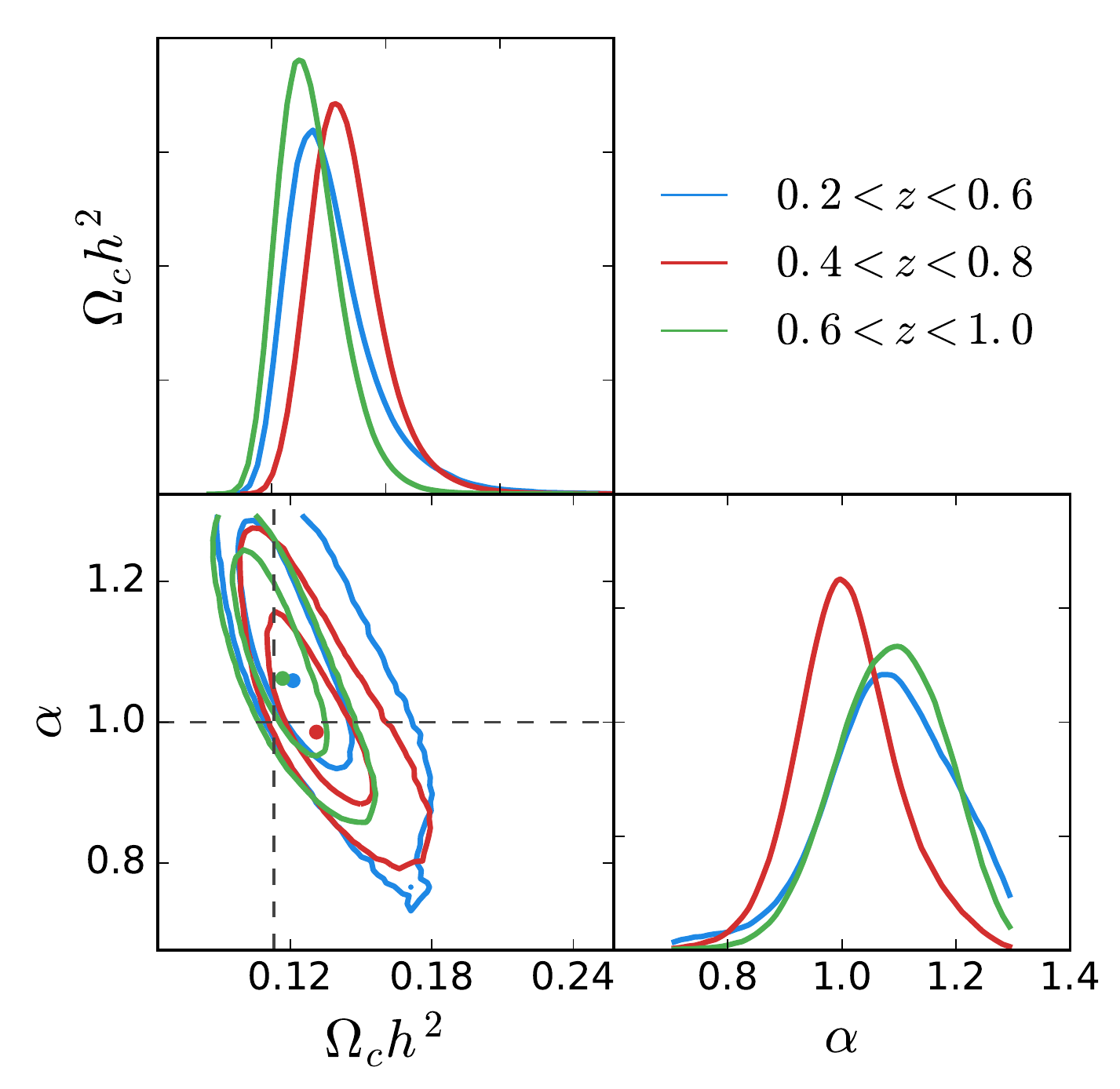}
	\end{center}
	\caption{Likelihood surfaces when fitting the 1D BAO signal from WiggleZ using the new WizCOLA covariance matrices. Parameters $b^2$, $\beta$, and $\sigma_V$ are marginalised over, with $\sigma_v$ set to $5 h^{-1}$ Mpc. The three redshift bins, $0.2<z<0.6$, $0.4<z<0.8$ and $0.6<z<1.0$ are shown in blue, red and green respectively. Dashed lines represent the values of fiducial cosmology.}
	\label{fig:fmonopole}
\end{figure}

We first fit using $\sigma_v$ as a free parameter, and finding $\sigma_v$ unconstrained, fix it to a specific value. We fix $\sigma_v = 5 h^{-1}$ Mpc, which is approximately its theoretically expected value.  This gives tighter constraints due to the fewer degrees of freedom, and does not bias results since the fit is insensitive to the value of this parameter (mean parameter deviation between fixing $\sigma_v$ and fitting for $\sigma_v$ was less than $0.05\sigma$). The likelihood surface for our fits can be seen in Figure \ref{fig:fmonopole}, and our results are compared in Table \ref{tab:blakekazintable}.

\begin{table*}
	\centering
	\caption{A comparison between the fits found in this analysis and those found in \citet{BlakeKazin2011}. These analyse the same data, but this analysis uses a slightly different model and an improved covariance matrix. The results given in \citet{BlakeKazin2011} use mean statistics, whilst we utilise maximum likelihood statistics, but the dominant difference comes from the improved covariance matrix. We convert our fit results from $\Omega_c h^2$ to $\Omega_m h^2$ for a direct comparison, using our fixed fiducial value of $\Omega_b h^2$.}
	\begin{tabular}{cc|ccc|ccc}
		\specialrule{.1em}{.05em}{.05em} 
		Sample & $z_{\rm{eff}}$ & \multicolumn{3}{c}{\citet{BlakeKazin2011}}  & \multicolumn{3}{c}{This analysis}\\
		&  & $\chi^2/{\rm DoF}$ & $\Omega_m h^2$ &$\alpha$ & $\chi^2/{\rm DoF}$ & $\Omega_m h^2$ & $\alpha$ \\
		\specialrule{.1em}{.05em}{.05em} 
		$0.2 < z < 0.6$ & $0.44$ & $0.88$ & $0.143\pm0.020$ &$1.024\pm0.093$ & $1.01$ & $0.143\pm 0.017$ & $1.07^{+0.13}_{-0.09}$ \\
		$0.4 < z < 0.8$ & $0.60$ & $0.78$ & $0.147\pm0.016$ &$1.003\pm0.065$ & $0.85$ & $0.151^{+0.017}_{-0.014}$ & $1.00^{+0.09}_{-0.08}$ \\
		$0.6 < z < 1.0$ & $0.73$ & $1.05$ & $0.120\pm0.013$ &$1.113\pm0.071$ & $1.22$ & $0.138^{+0.012}_{-0.015}$ & $1.10^{+0.09}_{-0.10}$ \\
		\specialrule{.1em}{.05em}{.05em} 
	\end{tabular} \label{tab:blakekazintable}
\end{table*}

We see that most results follow closely those obtained in \citet{BlakeKazin2011}, however the result of $\Omega_m h^2 (z_{\rm eff} = 0.73)$ shifts by more than $1\sigma$. This shift increases the value of $\Omega_m h^2 (z_{\rm eff} = 0.73)$, bringing it into agreement with the $\Omega_m h^2$ determination from the other redshift bins. These shifts were confirmed to be due to the change in covariance matrix and not any difference in changes to the modeling process, as rerunning the same fit using the original lognormal realisations bought the deviation to below $0.5\sigma$. Comparisons against fits using the WizCOLA covariance and lognormal covariance indicate in all redshift bins that the dominating contribution in fit difference is due to the improved covariance.

\subsection{Fitting range} \label{sec:trunc}
Before fitting our model correlation function to the data, we need to assess the range of scales over which the data are useful.  We expect the model to do poorly at small distances (where non-linearities are strong), while the data's sample variance increases at large distances.  So there will be an optimal range of scales over which to fit the data, and that range will be dependent on the data set (less biased tracers can go to smaller scales, larger volume data sets can go to larger scales).  We determined our optimal fitting range using simulations (see \S\ref{app:truncation})  and conclude that a data range of $25 < s < 180 \hMpc$ allows maximum utilisation of available data without introducing bias into our model.

\subsection{Validating the multipole expansion model against WizCOLA data}

 To validate our model we test it against the WizCOLA simulations.  WizCOLA used known survey geometry and an underlying fiducial cosmology to simulate 600 realisations of the WiggleZ survey, where the fiducial cosmology is parameterised by $\Omega_m = 0.273$, $\Omega_\Lambda = 0.727$, $\Omega_b = 0.0456$, $h=0.705$, $\sigma_8 = 0.812$ and $n_s = 0.961$ following WMAP cosmology \citep{Komatsu2009}.  Putting this in terms of $\Omega_c h^2$, we have $\Omega_c h^2 = 0.113$. We fit against these individual realisations and compare the distribution of our recovered results to the known fiducial cosmology, and also fit to the mean of all 600 simulations to create a single high quality dataset, where the standard deviation of the data was reduced by a factor of $\sqrt{600}$ as the simulations are independent.

As the cosmology used in the WizCOLA simulations is identical to the fiducial cosmology values used to extract data from the WizCOLA simulations, we do not expect to observe anisotropic warping when fitting to the simulation realisations.   As such, we can validate our model by ensuring that it recovers $\alpha = 1.0$ and $\epsilon = 0.0$ when fitting the WizCOLA correlation functions.

  Prior analyses have found poorly constrained values for $\sigma_v$ \citep{BlakeDavis2011} using WiggleZ data, and to validate that the bounds applied to $\sigma_v$ in prior analyses were not influencing or biasing fits, we take the mean realisation dataset (with its increased data strength) and fit to $\log(k_*)$ instead of $\sigma_v$, where the shift into log scale allows us to check values typically outside of allowed prior ranges. Thus we can confirm if the best-fitting $\log(k_*)$ value (and associated value of $\sigma_v$) fall within the predictions of current theory and past priors. Final parameter constraints are detailed in Table \ref{tab:wizmp}.

For all redshift bins, our best fits recovered the fiducial parameters well within the $1\sigma$ uncertainty limit. We can also see that, looking at the mean value of the determined values for $\log(k_*) = -2.10$, this gives a $\sigma_v = 5.77 h^{-1}$ Mpc, which is in the magnitude expected by the theory given in equation \eqref{eq:sigmav} and the values found in \citet{BlakeKazin2011} and \citet{BlakeDavis2011}.
Within the range $\sigma_v \in [0,10]$, we find no significant difference in $\chi^2$ values, indicating that $\sigma_v$ is not tightly constrained within theoretically predicted ranges. Fixing $\sigma_v = 5 h^{-1}$ Mpc, as we did in the 1D example, has negligible effect on the cosmological parameters of interest, and therefore we do that for the rest of our analysis.

\begin{table*}
	\centering
	\caption{Recovered parameter constraints when fitting to the combined 600 realisations of the WizCOLA simulation data multipoles, where the uncertainty given by the WizCOLA simulations has been reduced by a factor of $\sqrt{600}$ to account for the independent nature of each mock. Minimum $\chi^2$ values correspond to 39 DoF.}
	\begin{tabular}{cc|ccccc}
		\specialrule{.1em}{.05em}{.05em} 
		Sample & $z_{\rm{eff}}$ & min $\chi^2$ & $\Omega_c h^2$ &$\alpha$ & $\epsilon$ & $\log(k_*)$\\
		\specialrule{.1em}{.05em}{.05em} 
		$0.2 < z < 0.6$ & $0.44$ & $12.0$ & $0.112^{+0.005}_{-0.006}$ &$1.006^{+0.023}_{-0.022}$ & $0.002^{+0.016}_{-0.016}$ & $-2.18^{+0.22}_{-0.20}$\\
		$0.4 < z < 0.8$ & $0.60$ & $8.6$  & $0.114^{+0.005}_{-0.004}$ &$1.004^{+0.017}_{-0.016}$ & $0.005^{+0.012}_{-0.010}$ & $-2.05^{+0.20}_{-0.17}$\\
		$0.6 < z < 1.0$ & $0.73$ & $10.8$ & $0.113^{+0.005}_{-0.005}$ &$1.006^{+0.021}_{-0.018}$ & $0.007^{+0.015}_{-0.012}$ & $-2.07^{+0.26}_{-0.23}$\\
		\specialrule{.1em}{.05em}{.05em} 
		Input & & & 0.113 & 1.0 & 0.0 & \\
		\specialrule{.1em}{.05em}{.05em} 
	\end{tabular}\label{tab:wizmp}
\end{table*}

Some past surveys have included hexadecapole terms in the multipole analysis \citep{XuCuesta2013}. In order to test the significance of the hexadecapole term, we ran the above analysis with and without it. We find that the statistical uncertainty dominates any loss of information contained in the hexadecapole signal. Due to computational constraints and the low impact of the term, the hexadecapole contribution was left out of the final model.

\begin{figure}
	\begin{center}
		\includegraphics[width=\columnwidth]{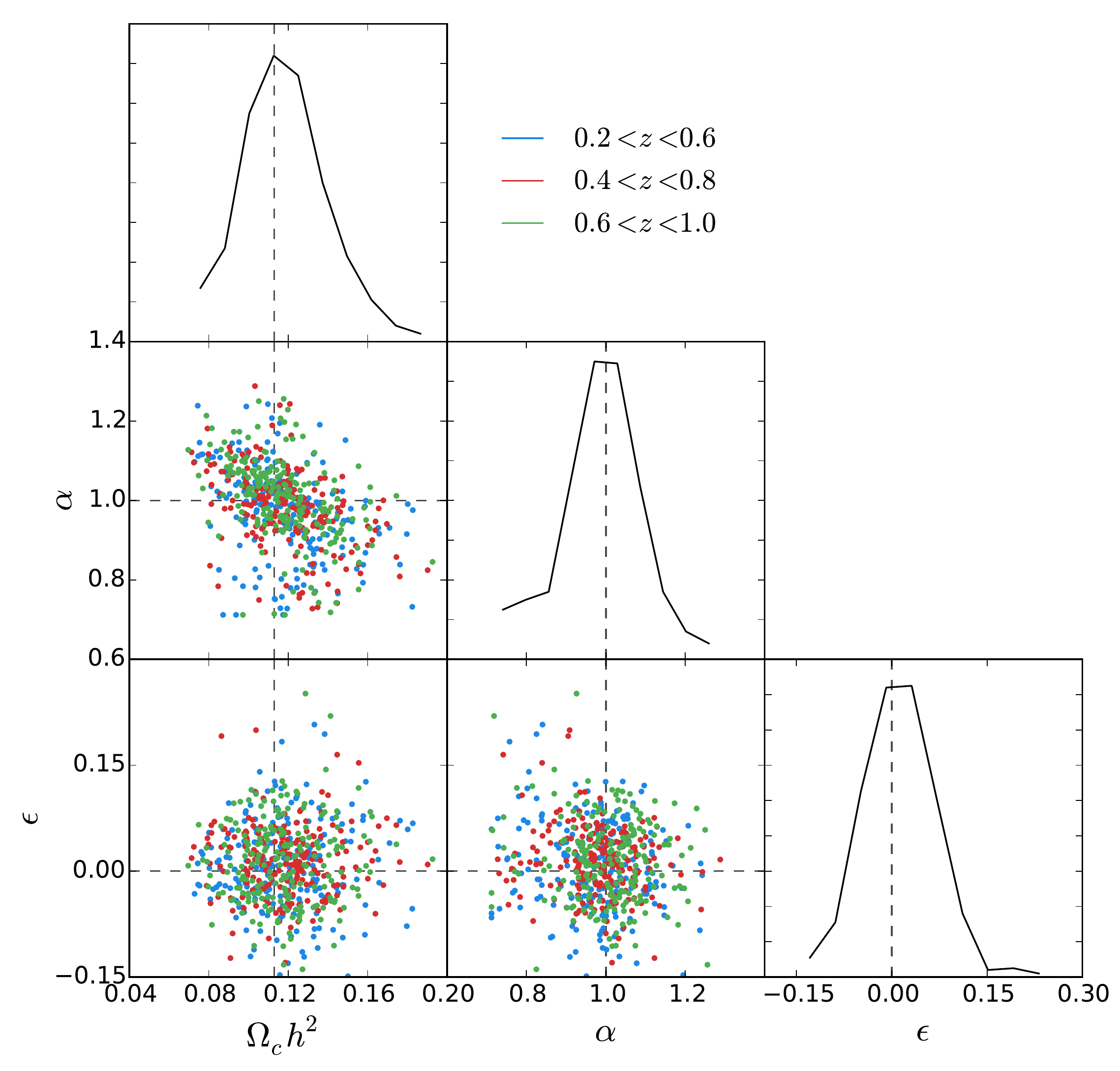}
	\end{center}
	\caption{Maximum likelihood $\Omega_c h^2$, $\alpha$ and $\epsilon$ values from WizCOLA realisations of the WiggleZ multipole data are shown in the bottom left corner plots. Dashed black lines indicate simulation parameters, and the solid black distributions in the diagonal subplots represent the final distribution across all bins for the specific parameter.}
	\label{fig:mpDist2}
\end{figure}

We can perform a validation of the multipole methodology by fitting to individual realisations of the WizCOLA simulation instead of the mean data set. 
The results are shown in Figure \ref{fig:mpDist2}, which confirms that the recovered parameter distribution matches the simulation.  

For small $\epsilon$, cosmological parameters can be extracted via Eq.~\ref{eq:alpha1} and Eq.~\ref{eq:alpha2}.

\begin{figure}
	\begin{center}
		\includegraphics[width=0.85\columnwidth]{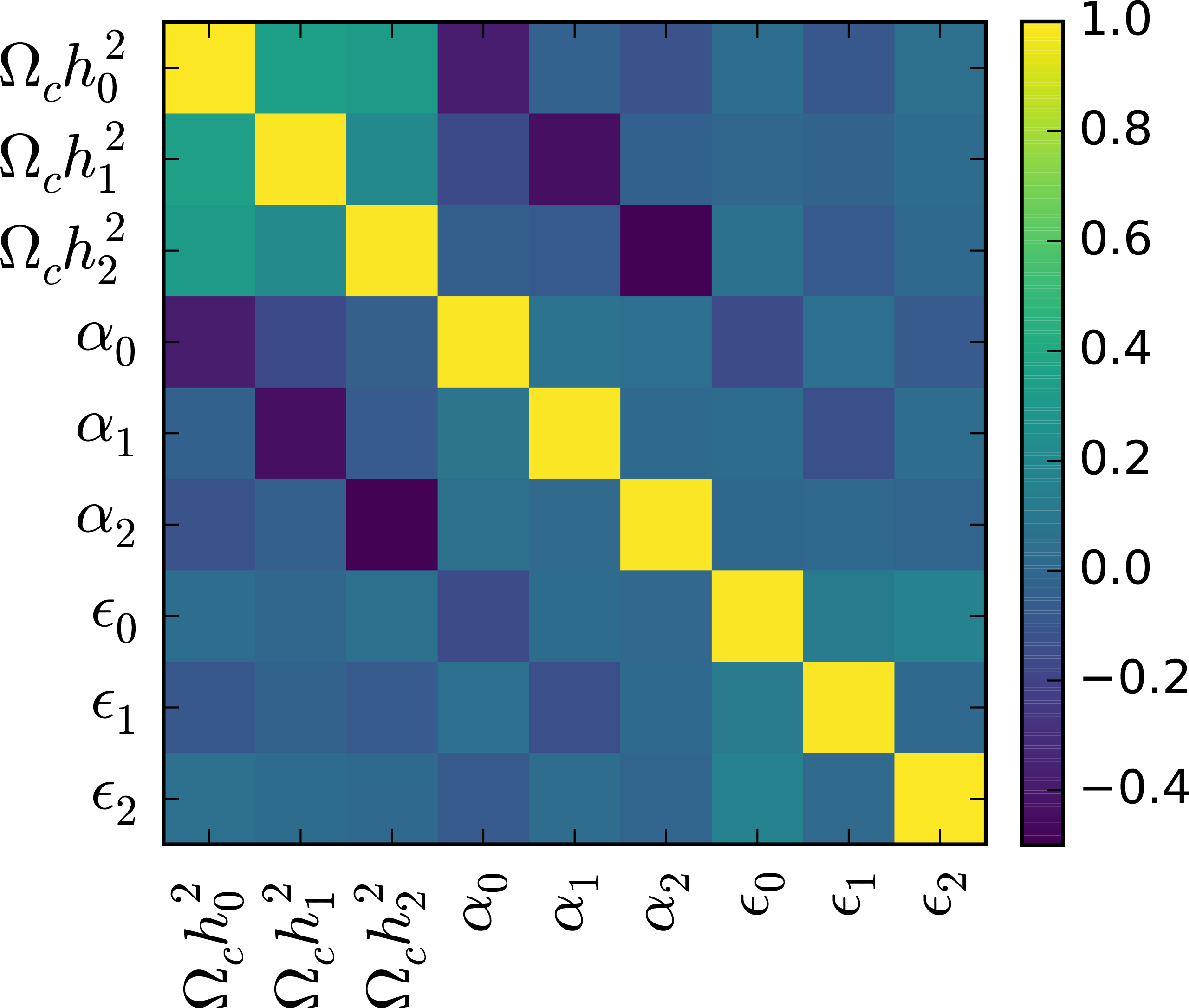}
	\end{center}
	\caption{Correlations between final cosmological parameters when fitting to the three redshift bins of each WizCOLA simulation for the multipole data. The subscript numbers after each parameter are used to denote the redshift bin, with $0$, $1$, and $2$ respectively denoting the $0.2 < z < 0.6$, $0.4 < z < 0.8$, and $0.6 < z < 1.0$ bins. }
	\label{fig:correlations}
\end{figure}

\begin{figure}
	\begin{center}
		\includegraphics[width=\columnwidth]{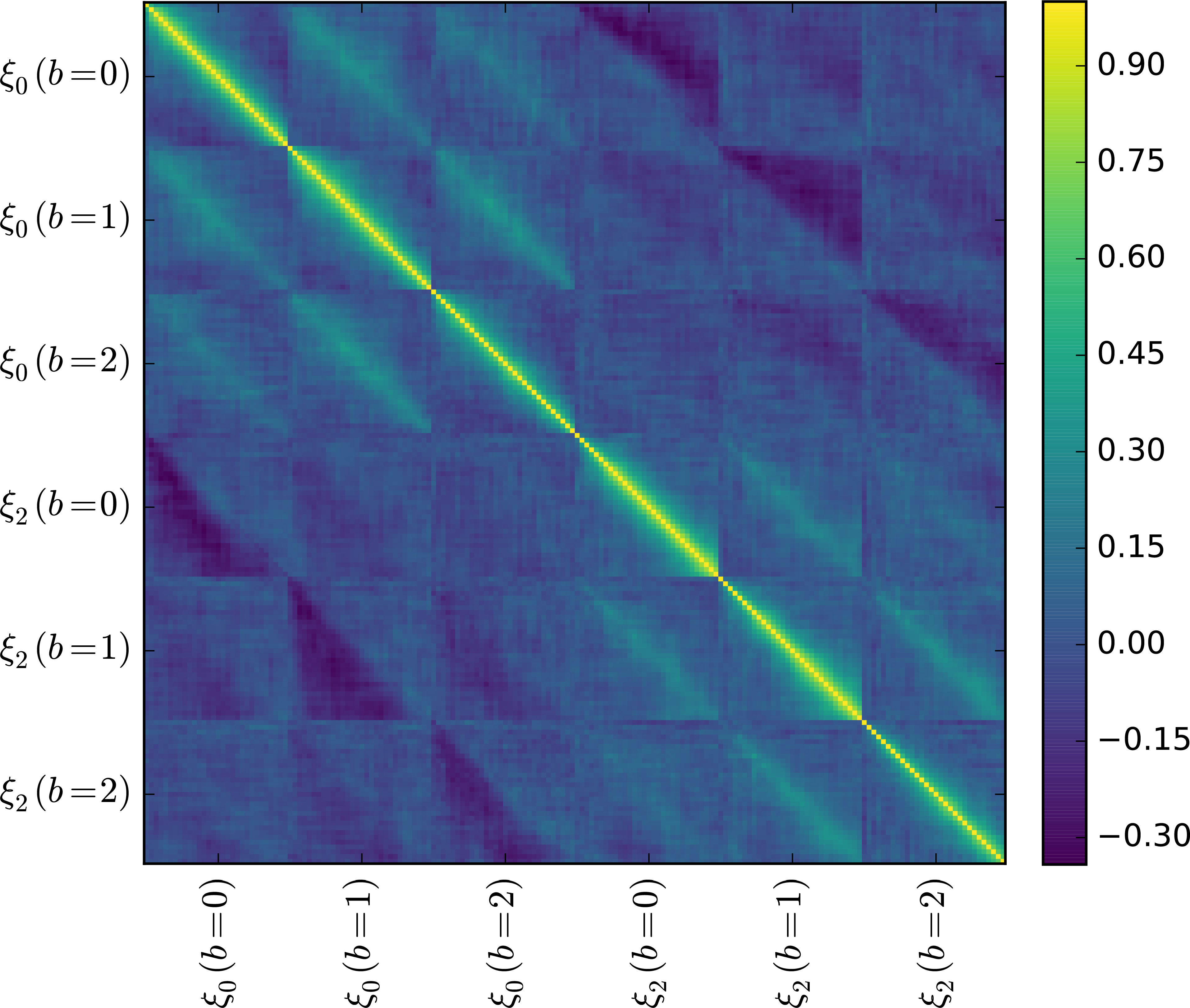}
	\end{center}
	\caption{Full data correlation matrices constructed for both the multipole expression of the WizCOLA data. The $b=0$, $b=1$ and $b=2$ labels respectively refer to the redshift bins $0.2 < z < 0.6$, $0.4 < z < 0.8$ and $0.6 < z < 1.0$. We can see that, even though the $b=0$ and $b=2$ bins do not overlap, some faint correlation still persists. This is expected, as data is generated using different snapshots of the same initial conditions, and thus the same set of modes are imprinted in both measurements.}
	\label{fig:fullCorrelations}
\end{figure}

\subsection{Combining redshift bins for multipole data}

The data present in the WizCOLA simulations and the final WiggleZ dataset is available in three redshift bins, $0.2 < z < 0.6$, $0.4 < z < 0.8$, and $0.6 < z < 1.0$. If these bins were independent, we could obtain our final parameter constraints  simply by combining the results for each individual bin. However, the redshift bins that we have chosen overlap and are thus correlated. 

There are two methods we can use to combine the binned data, and we utilise both methods in our multipole analysis so that we can check they give consistent results. The first method uses the correlation between final parameter values, and the second method calculates the covariance between the correlation function data points across all redshift bins and runs a separate fit that utilises all available data simultaneously.

\subsubsection{First Method: Parameter Covariance} \label{sec:parameterCov}

In order to determine final parametrisations across all redshift bins, the correlation between fit parameters from individual redshift bins needs to be quantified and accounted for. To do this, we fit individual realisations of the WizCOLA simulation, and construct a $9\times 9$ covariance matrix from the peak likelihood fit values for parameters ($\Omega_c h^2$, $\alpha$ and $\epsilon$ for a multipole analysis), such that we construct,
\begin{align}
C_{ij} = \frac{1}{N-1} \sum\limits_{n=1}^{N} (\theta_{i,n} - \bar{\theta}_i)(\theta_{j,n} - \bar{\theta}_j),
\end{align}
where $\theta$ represents the list of parameters, such that $\theta_{i,n}$ represents the value of $\theta_i$ on the $n$th WizCOLA realisation.
\begin{align*}
\theta=\left\lbrace \Omega_c h^2 (z = 0.44), \Omega_c h^2 (z = 0.60),  \Omega_c h^2 (z = 0.73), \right. \\ 
\alpha (z = 0.44), \alpha (z = 0.60),  \alpha (z = 0.73), \\
\epsilon (z = 0.44), \left. \epsilon (z = 0.60),  \epsilon (z = 0.73) \right\rbrace.
\end{align*}
Similarly to the covariance matrix, we can also calculate the correlation matrix, defined as
\begin{align}
R_{ij} = \frac{1}{N-1} \sum\limits_{n=1}^{N} \frac{(\theta_{i,n} - \bar{\theta}_i)(\theta_{j,n} - \bar{\theta}_{j})}{\sigma_i \sigma_j},
\end{align}
where $\sigma_i$ represents the standard deviation of the $i$th parameter. The correlation matrix $R_{ij}$ determined from analysis of the WizCOLA realisations is shown in Figure \ref{fig:correlations}.

This covariance matrix can now be used to fit for a final $\Omega_c h^2$, $\alpha$ and $\epsilon$, by minimising the $\chi^2$ statistic, given as,
\begin{align} \label{eq:covchi}
\chi^2(\Omega_c h^2, \alpha, \epsilon) &= (\Omega_c h^2 - \Omega_c h^2_0, \Omega_c h^2 - \Omega_c h^2_1, ...,\  \epsilon - \epsilon_2)^T \notag \\
 &C_{ij}^{-1}(\Omega_c h^2 - \Omega_c h^2_0, \Omega_c h^2 - \Omega_c h^2_1, ...,\  \epsilon - \epsilon_2),
\end{align}
where again the subscript indices on the $\Omega_c h^2$, $\alpha$ and $\epsilon$ refer to the redshift bins. In essence, we utilise the parameters fitted to each bin as datapoints in a secondary model, which we minimise with respect to the final parameters $\Omega_c h^2$, $\alpha$ and $\epsilon$.

\subsubsection{Second method: All data covariance} \label{sec:allData}

The covariance matrices utilised so far in our analysis have been supplied from the WizCOLA simulations, and give data covariance inside each redshift bin. However, also having the 600 WizCOLA realisations, we can reconstruct a full covariance matrix to give the covariance between values of the correlation function across redshift bins. The correlation matrices for the multipole data are shown in Figure \ref{fig:fullCorrelations}.

\begin{figure}
	\begin{center}
		\includegraphics[width=\columnwidth]{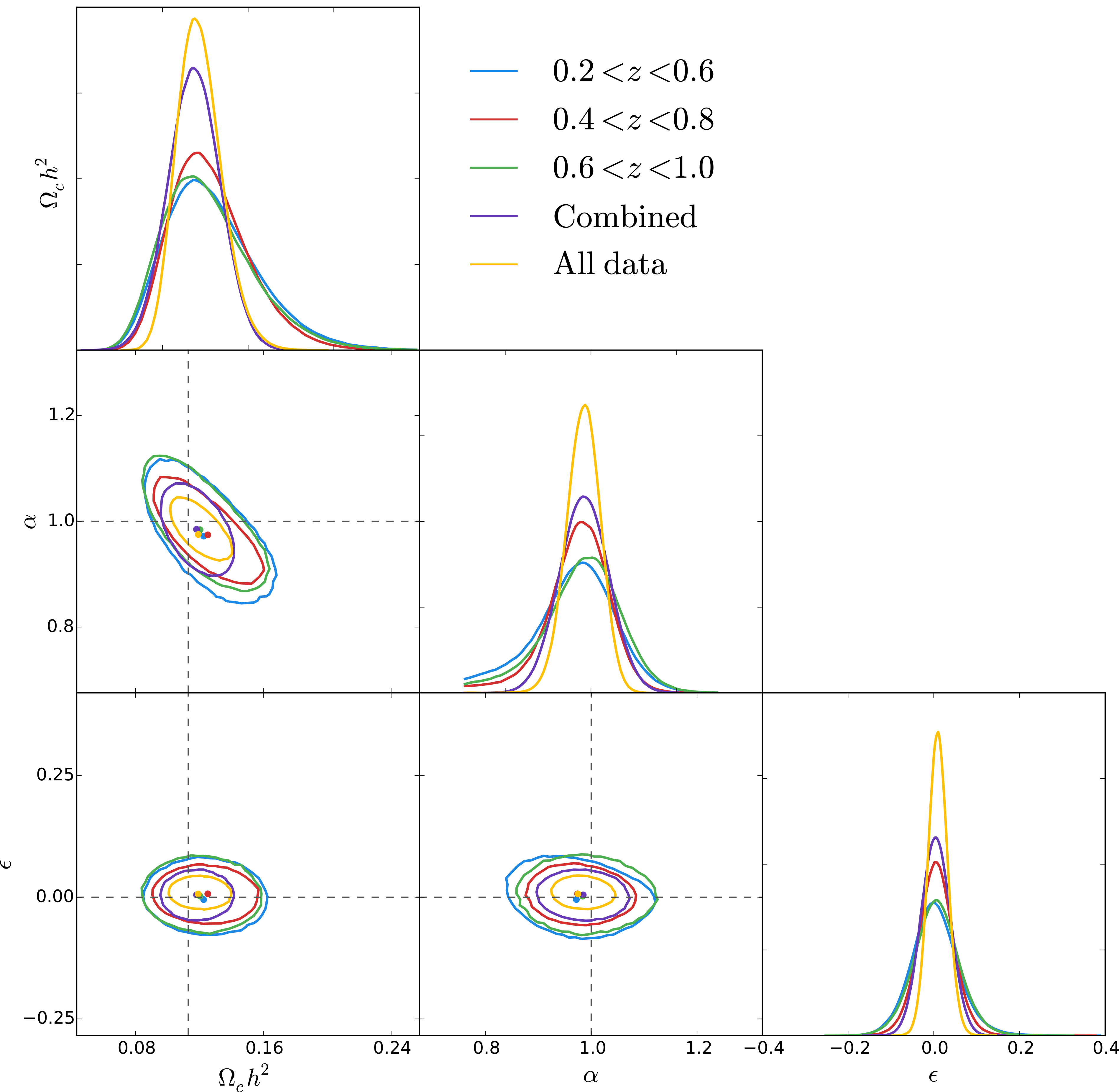}
	\end{center}
	\caption{Fits to the mean data of all 600 WizCOLA realisations 
		for the multipole expansion expression of the data. Fits using all three bins simultaneously are shown as the ``All data'' fits, and the combination of maximum likelihood parameters from 3 bins using parameter covariance is shown as the ``Combined'' likelihood surfaces. In all cases we recover simulation cosmology well within $1\sigma$ limits. The three redshift bins, $0.2<z<0.6$, $0.4<z<0.8$ and $0.6<z<1.0$ are shown in blue, red and green respectively. Dashed lines represent the values of fiducial cosmology.}
	\label{fig:corCombinedMPWiz}
\end{figure}

When using the full data covariance to simultaneously fit all three redshift bins, a further question becomes whether marginalisation parameters $b^2$, $\beta$, $\sigma_v$, and $\sigma_V$ should be free between redshift bins, or consistent across them. 

From a physical motivation, we expect the bias parameter $b^2$ to be dependent on redshift bin. This is because we only observe the most massive, luminous galaxies at high redshift, which have higher bias than the less massive galaxies we can see at lower redshifts. However, when performing fits, $b^2$ and $\beta$ are well constrained, whilst $\sigma_v$ and $\sigma_V$ are not. As this implies that those two parameters do not significantly contribute to the likelihood calculations, it is unknown if setting $\sigma_V$ free between bins will have a noticeable benefit.

To investigate this, we ran fits to the combined WizCOLA data where we set no nuisance parameters free between redshift bins, when we only set $b^2$ free, when we set all \textit{but} $b^2$ free, and then when we set all four nuisance parameters free.  These fits indicate a strong preference for fitting with separate $b^2$ values in different redshift bins due to tighter constraints achieved, and an accompanying small improvement in $\chi^2$.  However allowing the other nuisance parameters to vary between redshift bins has negligible benefits (it neither decreases the uncertainty in parameter fits nor removes  bias), and adds computational time in the form of delayed chain convergence.

Based on these results, we utilise independent $b^2$ values, whilst fixing $\beta$, $\sigma_v$, and $\sigma_V$ between bins when fitting with the full data set and full data covariance.

\subsection{Multipole model testing conclusions}

 A graphical comparison of fits to the mean WizCOLA data for individual redshift bins, all data fits, and combining bin parameters can be found for the multipole data format in Figure \ref{fig:corCombinedMPWiz}. 
 The results are consistent between bins, and between methods of combining bins, and all are consistent with the input cosmology.  We therefore conclude that our model can accurately be used to derive cosmological constraints from WiggleZ-like data.  The results with real data are presented in Section~\ref{sec:multi}, but before presenting these results we continue our validation testing, now on the reconstructed data.

\section{Validating reconstructed wedge data}

\begin{figure}
\begin{center}
\includegraphics[width=\columnwidth]{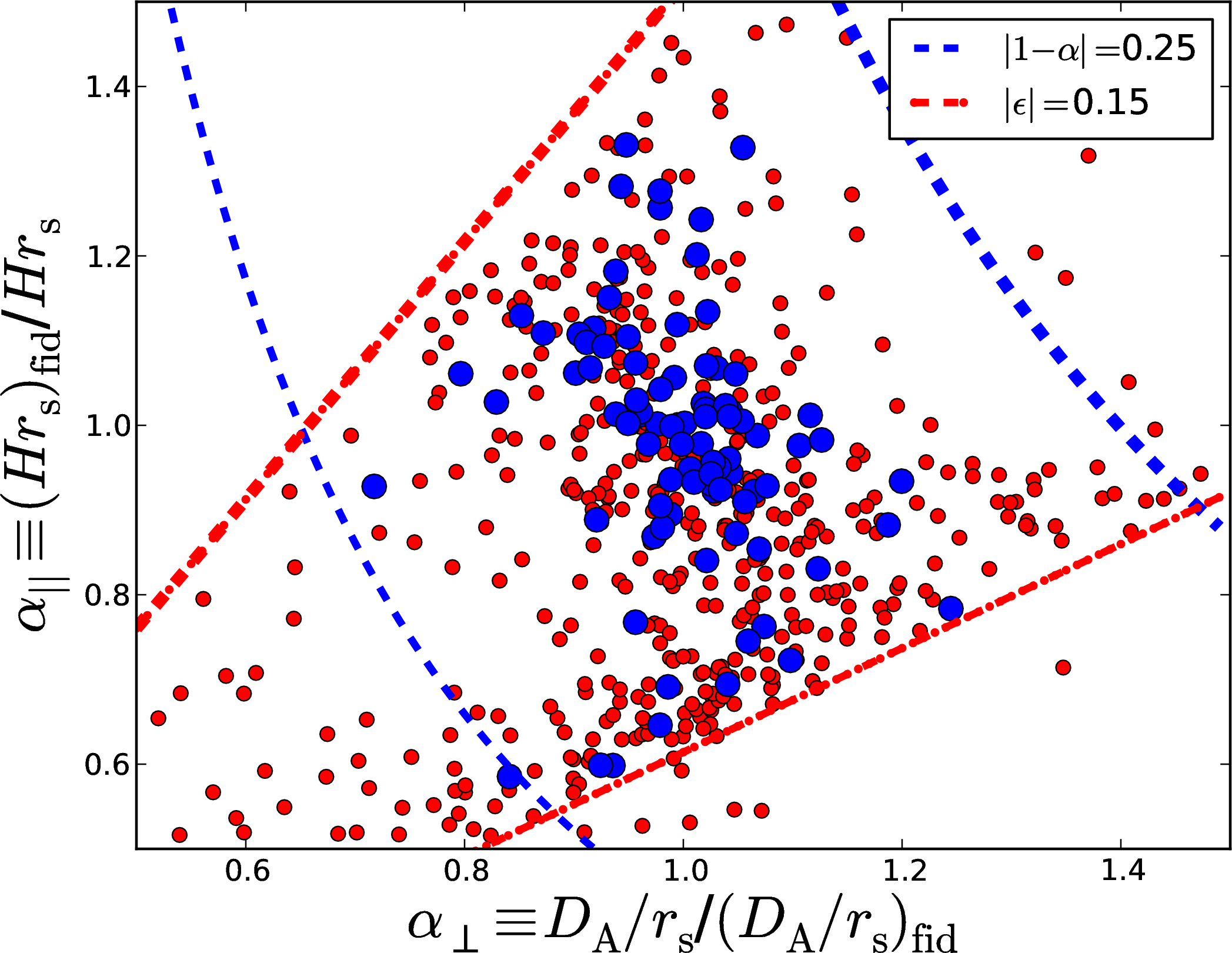}
\caption{\label{fig:wizcola_hdaModes_z60_epsilonT15} WizCOLA simulation fits for the redshift $0.6<z<1$ bin using the wedge data. The large blue circles (87/600) are realizations that have a significance of detection of 2.9$\sigma$ (as the observation) or higher. The red circles are below this threshold. The red dot-dashed lines mark the $\pm15\%$ value of the fiducial $\epsilon$ which we use as a flat prior in this calculation. The dashed blue lines mark $\pm25\%$ of the fiducial $\alpha$, but are just for visualisation as we did not apply these as a prior. The thicker lines indicate the higher values of the $\alpha$ and $\epsilon$ limits.%
}
\end{center}
\end{figure}

\begin{figure}
\begin{center}
\includegraphics[width=\columnwidth]{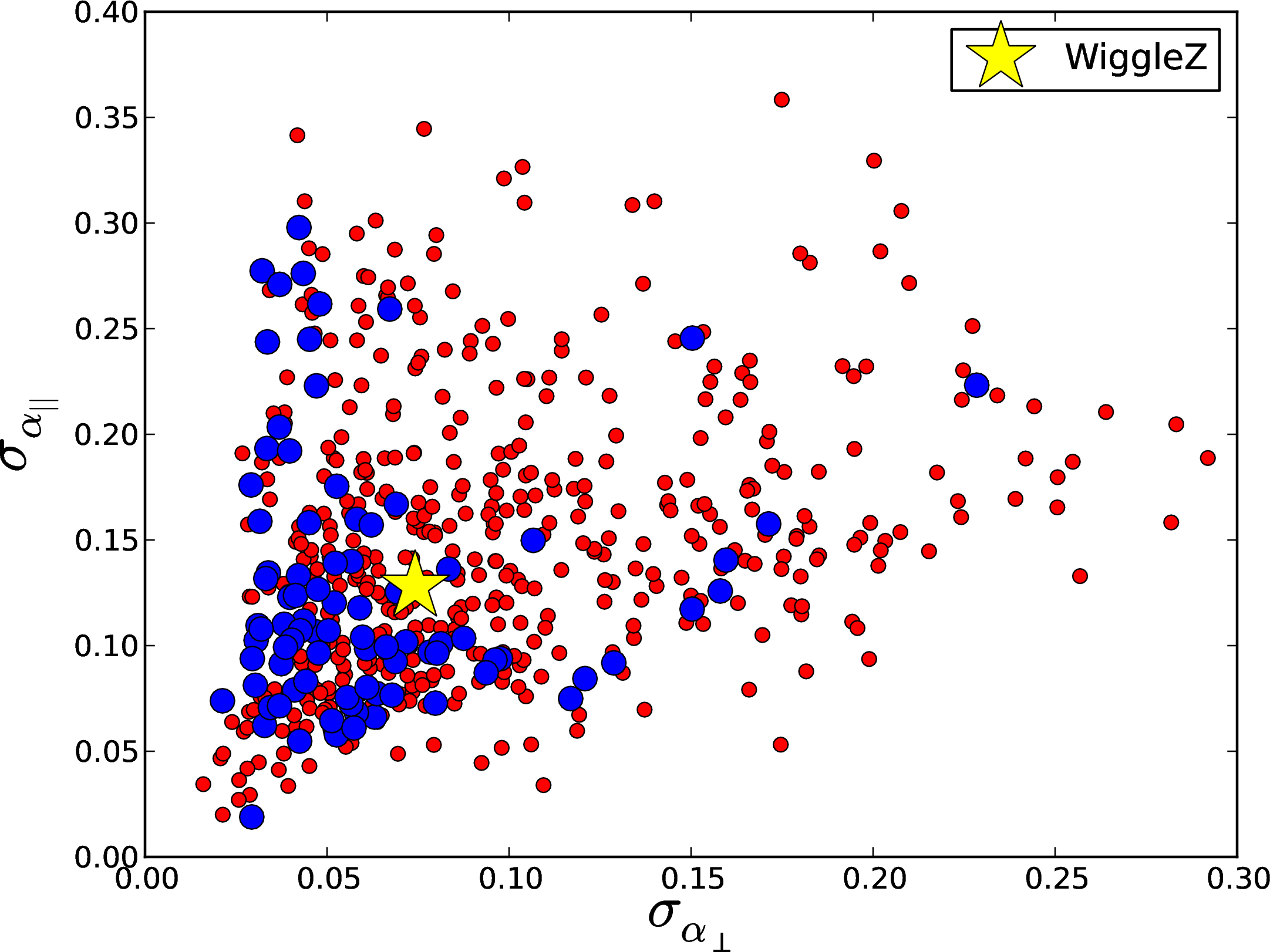}
\caption{\label{fig:wizcola_hdaUnc_z60_epsilonT15} Uncertainty on the WizCOLA simulation fits for the redshift $0.6<z<1$ bin using the wedge data. The large blue circles are realizations that have a significance of detection of 2.9$\sigma$ (same as the observation) or higher. The red circles are below this threshold. For comparison, the star is our WiggleZ result.%
}
\end{center}
\end{figure}

For the reconstructed data that we analyse in wedges we also tested our procedure using the WizCOLA mock catalogues.  We focus
here on results from the mocks of the $\Delta z^{\rm Far}$ redshift
slice, which are representative of the behaviour in all redshift bins.
In Figure \ref{fig:wizcola_hdaModes_z60_epsilonT15} we present the
best-fitting values of $\alpha_\parallel$ and $\alpha_\perp$ from each
of the 600 simulations, and in Figure
\ref{fig:wizcola_hdaUnc_z60_epsilonT15} we show the corresponding
uncertainties. Post-reconstruction data fitting ranges follow \citet{KazinKoda2014}, with bin 
separations of $6.7 h^{-1}\, {\rm Mpc}$ and fitting range $s > 50 h^{-1}\,{\rm Mpc}$. 
The uncertainties are typically large, indicating a
marginal detection of the baryon acoustic peak in the clustering
wedges.  This motivated us to consider, in addition to the $50\%$
priors on $\alpha_\parallel$ and $\alpha_\perp$ mentioned above,
additional flat priors on their combination, which we parameterize as
$\alpha = \alpha_\perp^{2/3} \alpha_\parallel^{1/3} \propto D_{\rm
  A}^2 r^\prime_s /(H r_s)$ and $\epsilon = (\alpha_\parallel/\alpha_\perp)^{1/3} - 1
\propto 1/(D_{\rm A} H)$.
The $\alpha$ parameter is mostly sensitive to
the monopole and $\epsilon$ to the quadrupole, although both terms appear in all
multipoles \citep[see][for a discussion]{PadmanabhanWhite2008}. In the
final analysis we applied a $15\%$ flat prior on $\epsilon$, which is
marked by the red dot-dashed lines in Figure
\ref{fig:wizcola_hdaModes_z60_epsilonT15}.  We did not apply a prior
in $\alpha$, but for illustration we show the $\pm 25\%$ threshold as
the blue dashed lines in Figure
\ref{fig:wizcola_hdaModes_z60_epsilonT15}.  We selected from the 600
realizations those that have a significance of BAO detection equal to
or greater than that in the real dataset ($2.9\sigma$).  We find 87
such mocks (15\%; marked as large blue circles).  In Figure
\ref{fig:wizcola_hdaUnc_z60_epsilonT15} we also show our WiggleZ
$\Delta z^{\rm Far}$ result with a yellow star.

Many WiggleZ mock realizations do not permit good constraints on both
$\alpha_\perp$ and $\alpha_\parallel$.  However, for the subset of
realizations with similar detection significance to the WiggleZ data,
we find that our procedure enables us to extract unbiased distance
measurements, with median and standard deviations $\langle \alpha_\perp \rangle =
1.001 \pm 0.081$ and $\langle \alpha_\parallel \rangle = 1.00 \pm 0.15$.  The
median and standard deviation of the errors in these parameters, for
this subset of mocks, are $<\sigma_{\alpha_\perp}> = 0.052 \pm 0.037$
and $\langle \sigma_{\alpha_\parallel} \rangle = 0.107 \pm 0.061$.  Similar results
are obtained when analyzing mocks at $\Delta z^{\rm Mid}$ and $\Delta
z^{\rm Near}$.  In all cases, the results for the WiggleZ data are
consistent with the range covered by the simulations.

We now consider the degree to which our $15\%$ prior in $\epsilon$
impacts the model-independence of our results.  Using MCMC chains
based on {\sl Planck} temperature and {\sl WMAP} polarization data, we
found that the scatter in $\epsilon$ at our redshifts of interest was
$2.0\%$ for a flat $\Lambda$CDM model and $2.5\%$ for an $ow$CDM
model.  We hence argue that our much larger $15\%$ prior does not
significantly compromise our model-independence.

\begin{figure}
	\begin{center}
		\includegraphics[width=\columnwidth]{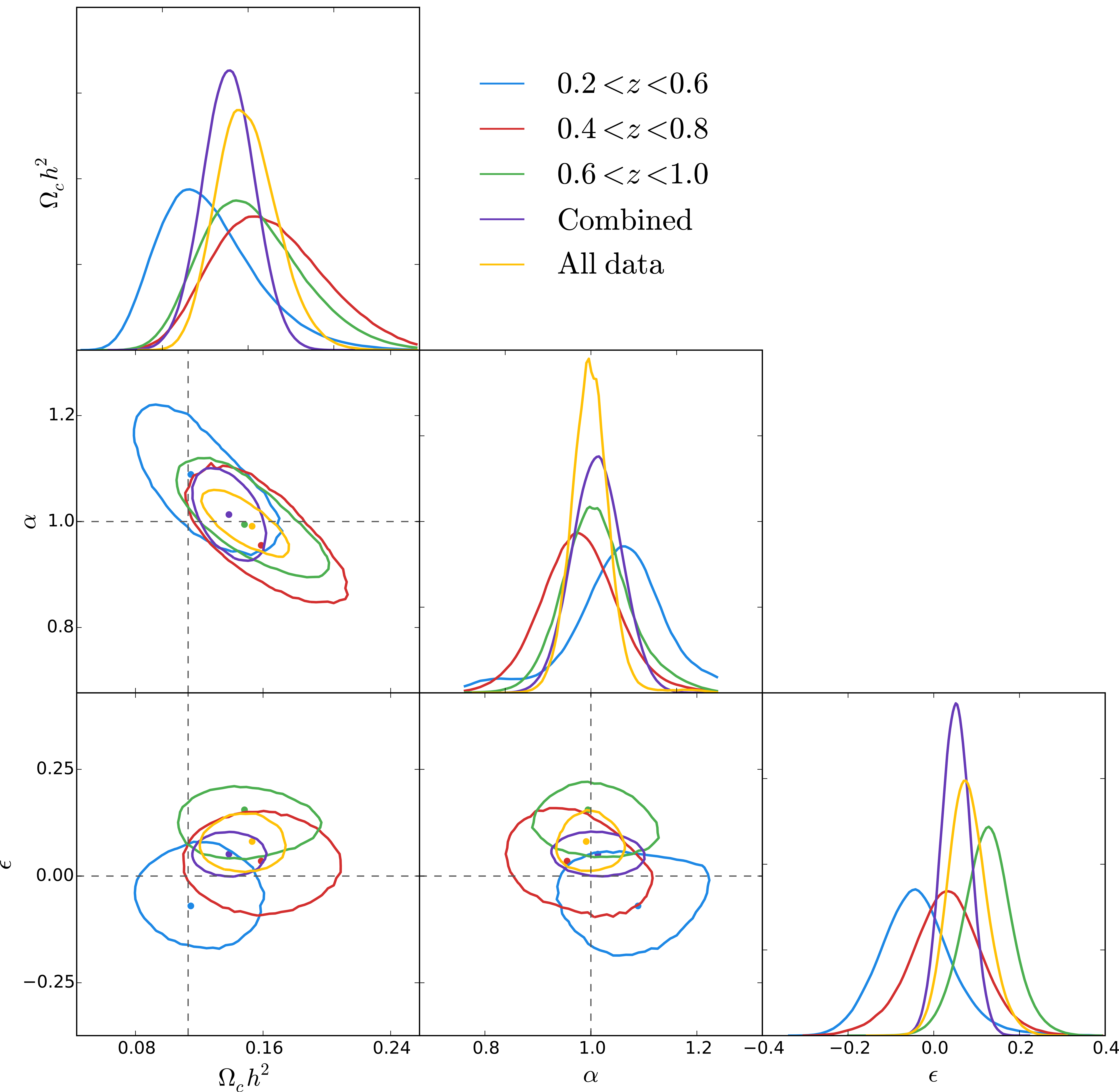}
	\end{center}
	\caption{Likelihood surfaces and marginalised distributions of $\Omega_ch^2$, $\alpha$ and $\epsilon$ for the WiggleZ multipole expression of the data. The three redshift bins, $0.2<z<0.6$, $0.4<z<0.8$ and $0.6<z<1.0$ are shown in blue, red and green respectively. Combining the fits of these three bins is shown as the purple ``Combined'' surface, and fitting for all the data simultaneously is shown in yellow. Dashed lines represent the values of fiducial cosmology.}
	\label{fig:wigglezBinsMP}
\end{figure}

\begin{figure}
	\begin{center}
		\includegraphics[width=\columnwidth]{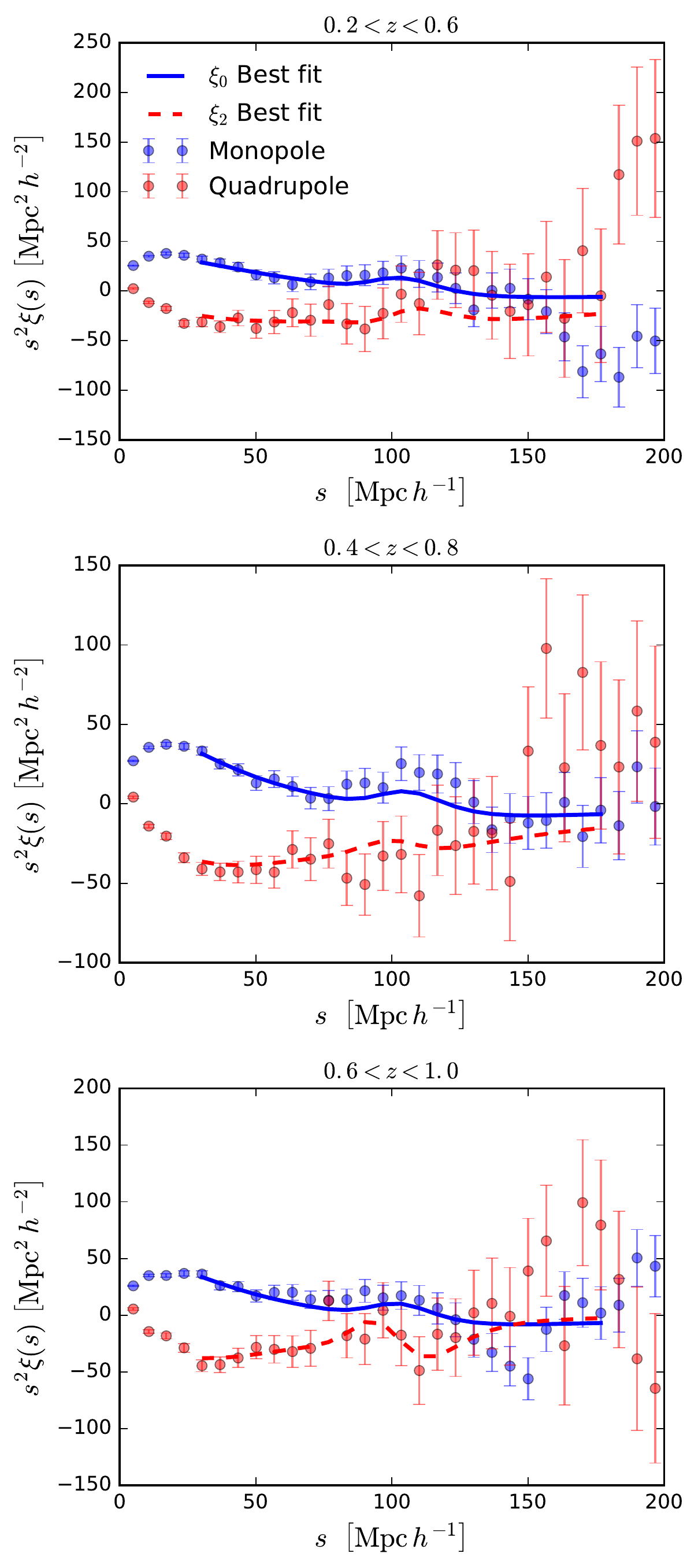}
		\caption{\label{fig:prerecon_result}  WiggleZ pre-reconstruction $0.2<z<0.6$ (upper), $0.4<z<0.8$ (mid), $0.6<z<1$ (lower) monopole $\xi_{0}$ (blue), quadrupole $\xi_{2}$ (red) and best-fit models.%
		}
	\end{center}
\end{figure}

\section{Unreconstructed Multipole Results}
\label{sec:multi}

Using the methodology outlined in Section~\ref{sec:test} we fit to the final unreconstructed WiggleZ dataset from \citet{KazinKoda2014} -- firstly fitting in each individual redshift bin and combining the results (Sect.~\ref{sec:parameterCov}), and secondly fitting all redshift bins simultaneously (Sect.~\ref{sec:allData}). The final distributions are given in Table \ref{tab:wigglezBinsParams} and illustrated in Figure~\ref{fig:wigglezBinsMP}.  The two methods give consistent results.

The conversion from $\alpha$ and $\epsilon$ to $D_A(z)$ and $H(z) $ is given by equations \eqref{eq:alpha1} and \eqref{eq:alpha2}. Using these relationships, we formulate parameter constraints. Figure \ref{fig:prerecon_result} displays the WiggleZ monopole and quadrupole data with the best fitting model overplotted in the three redshift ranges investigated. Baryonic acoustic peak signatures are present in all three redshift bins.

To determine the significance of the BAO peak detected in our analysis, we reran the multipole analysis with a model devoid of the BAO peak and converted the $\Delta \chi^2$ into a detection significance, which we found to be just over $2\sigma$ in all redshift bins. The low significance of the BAO peak is expected: the 1D BAO analysis from \citet{BlakeDavis2011} found a significance of $3.2\sigma$ when using all data in one combined bin, while our analysis used the data divided over three bins, and includes extra parameters to model angular dependence, so it is expected the statistical significance of the BAO peak would decrease. The analysis in \citet{BlakeKazin2011}, which utilised three redshift bins, the same as our analysis, found statistical significances between $1.9\sigma$ and $2.4\sigma$, consistent with our results.

\begin{table*}
	\centering
	\caption{Final parameter constraints from fitting the 2D BAO signal in the pre-reconstruction WiggleZ multipole correlation function. Minimum $\chi^2$ values correspond to 39 DoF. $D_A(z)$ given in units of Mpc, and $H(z)$ is presented in ${\rm km}\,{\rm s}^{-1} \, {\rm Mpc}^{-1}$. Correlation values are given in Appendix \ref{app:cor}.}
	\begin{tabular}{cc|cc|cccc|ccc}
		\specialrule{.1em}{.05em}{.05em} 
		Sample & $z_{\rm{eff}}$ & $D^\prime_A(z)$ & $H^\prime(z)$ &  $\chi^2$ & $\Omega_c h^2$  &$\alpha$ & $\epsilon$ & $D_A(z)$ & $H(z)$ & BAO peak significance\\
		\specialrule{.1em}{.05em}{.05em} 
		$0.2 < z < 0.6$ &  $0.44$ & 1175.5  & 87.4 & 55.0  & $0.117^{+0.029}_{-0.023}$ & $1.07^{+0.10}_{-0.10}$ & $-0.03^{+0.07}_{-0.10}$ & $1330 \pm  150$ & $85^{+19}_{-12}$  & $2.2\sigma$\\
		$0.4 < z < 0.8$ &  $0.60$  & 1386.2  & 95.5  & 69.3  & $0.156^{+0.035}_{-0.028}$ & $0.98^{+0.08}_{-0.10}$ & $0.05^{+0.07}_{-0.10}$ & $1280^{+190}_{-160}$ & $91^{+15}_{-14}$  & $2.1\sigma$\\
		$0.6 < z < 1.0$ &  $0.73$ & 1509.4  & 102.8 & 59.1 & $0.143^{+0.033}_{-0.026}$ & $1.00^{+0.08}_{-0.07}$ & $0.12^{+0.06}_{-0.05}$ & $1340^{+150}_{-130}$ & $80^{+9}_{-10}$  & $2.3\sigma$\\
		\specialrule{.1em}{.05em}{.05em} 
	\end{tabular}\label{tab:wigglezBinsParams}
\end{table*}

\begin{table*}
\begin{centering}
\caption{Model-independent measurements using the post-reconstruction $\xi_{||,\perp}$. The values quoted are the modes and the $68\%$ CL regions. The percentages indicate half of the $68\%$ CL regions. $r$ denotes the correlation between parameter fits. Values for the $z=0.44$ bin are not reported as the data was insufficient to provide constraints and final surfaces were heavily dependent on choice of prior.In the $\chi^2$ fitting we use 36 DoF. These results are displayed in Figure \ref{fig:hubble_diagram}.}
\label{tab:hda_wigglez_reconstructed}
\begin{tabular}{ l | r | r | c | c | c | c | c }
\hline
              & \multicolumn{2}{c|}{Fiducial} & \multicolumn{5}{c}{Measured} \\  
$z_{\rm eff}$ & $cz/(H r_{\rm s})$ & $D_{\rm A}/r_{\rm s}$  & $cz/(H r_{\rm s})$ & $D_{\rm A}/r_{\rm s}$  & $r$ & $\chi^2$ & BAO peak significance \\
\hline
  0.60  & 12.27 & 9.03 & 11.5$^{+1.3}_{-1.6}$ (13\%)  & 10.3$^{+0.4}_{-0.5}$ (5\%)   & -0.16  & 25 & $2.7\sigma$ \\
  0.73  & 13.87 & 9.84 & 15.3$^{+2.1}_{-1.8}$ (13\%)  &  9.8$^{+1.1}_{-0.4}$ (7\%)   & -0.36  & 35 & $2.9\sigma$ \\
\end{tabular}

\end{centering}
\end{table*}

\begin{figure}
	\begin{center}
		\includegraphics[width=0.9\columnwidth]{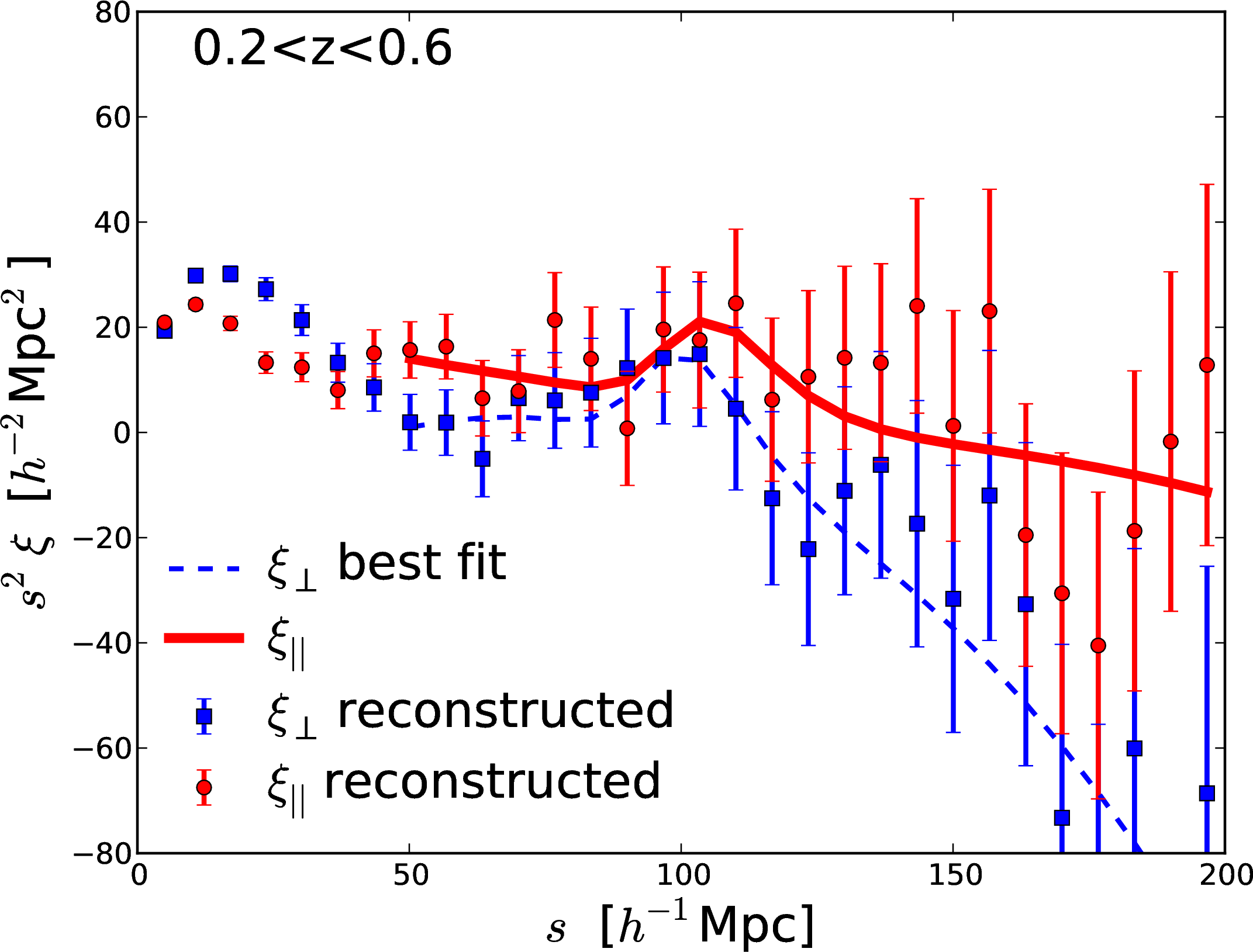}
		\includegraphics[width=0.9\columnwidth]{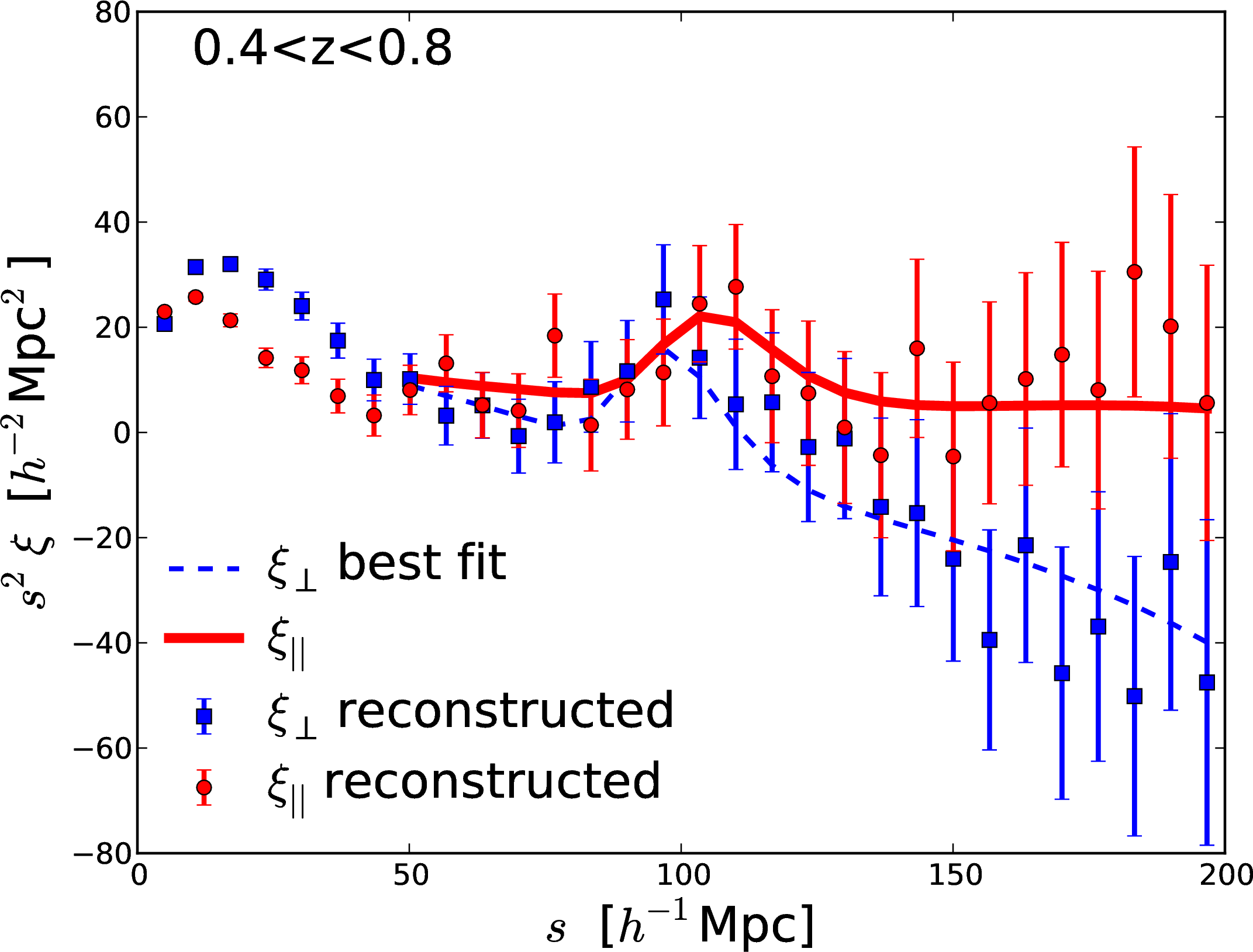}
		\includegraphics[width=0.9\columnwidth]{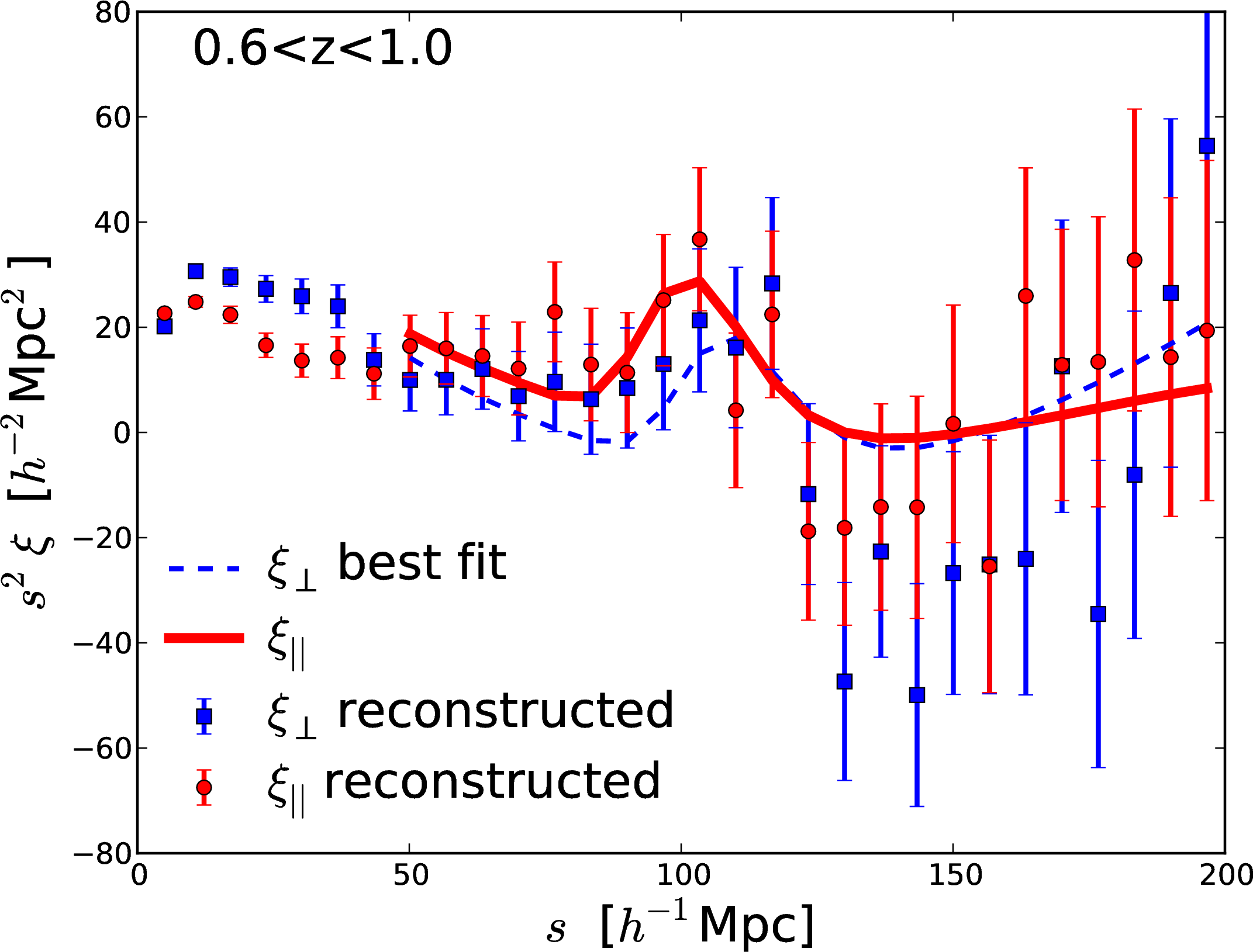}
		\caption{\label{fig:wigglez_wedges_z60}  WiggleZ post-reconstruction $0.2<z<0.6$ (upper), $0.4<z<0.8$ (mid), $0.6<z<1$ (lower) clustering $\xi_{||}$ (line-of-sight; red circles), $\xi_{\perp}$ (transverse; blue squares) and best-fit models.%
		}
	\end{center}
\end{figure}

\section{Reconstructed Results}\label{sec:wedge}

Figure \ref{fig:wigglez_wedges_z60} displays the clustering wedges
$\xi_\perp(s)$ (transverse wedge $\mu<0.5$; blue squares) and
$\xi_\parallel(s)$ (line-of-sight wedge $\mu>0.5$; red circles) in the
three redshift ranges investigated $\Delta z^{\rm Near}$, $\Delta
z^{\rm Mid}$, and $\Delta z^{\rm Far}$.  We overplot best-fitting
models for which we calculated $\chi^2 = 35.3, 24.8$, and $34.4$,
respectively, with 36 degrees of freedom.  We see baryonic acoustic
peak signatures in both $\xi_\perp$ and $\xi_\parallel$ for the $z=0.60$ and
$z=0.73$ redshift bins.  The fluctuations from zero at large scales are
consistent with the characteristic sample variance seen in the WizCOLA
simulations.

\begin{figure}
\begin{center}
\includegraphics[width=0.9\columnwidth]{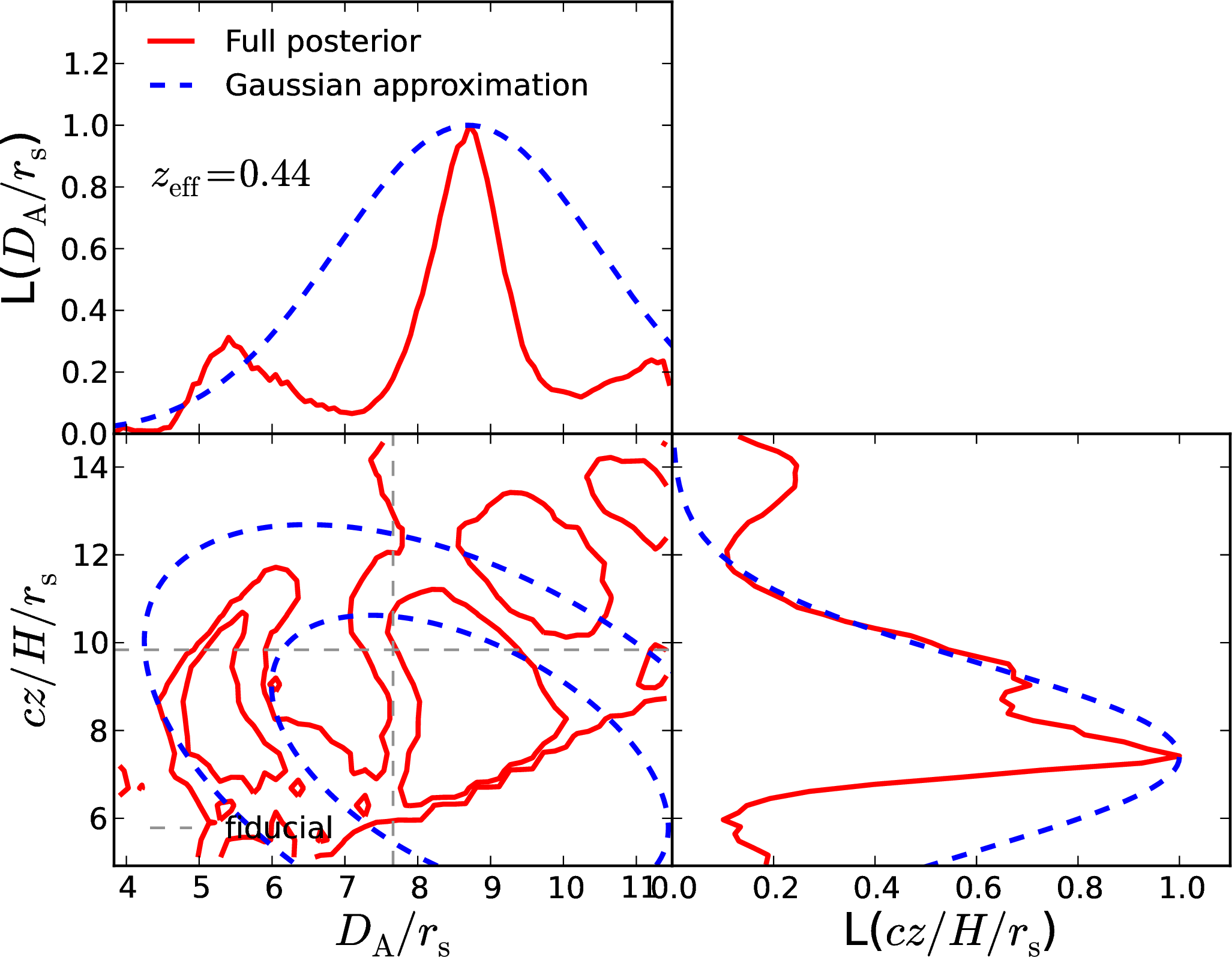}
\includegraphics[width=0.9\columnwidth]{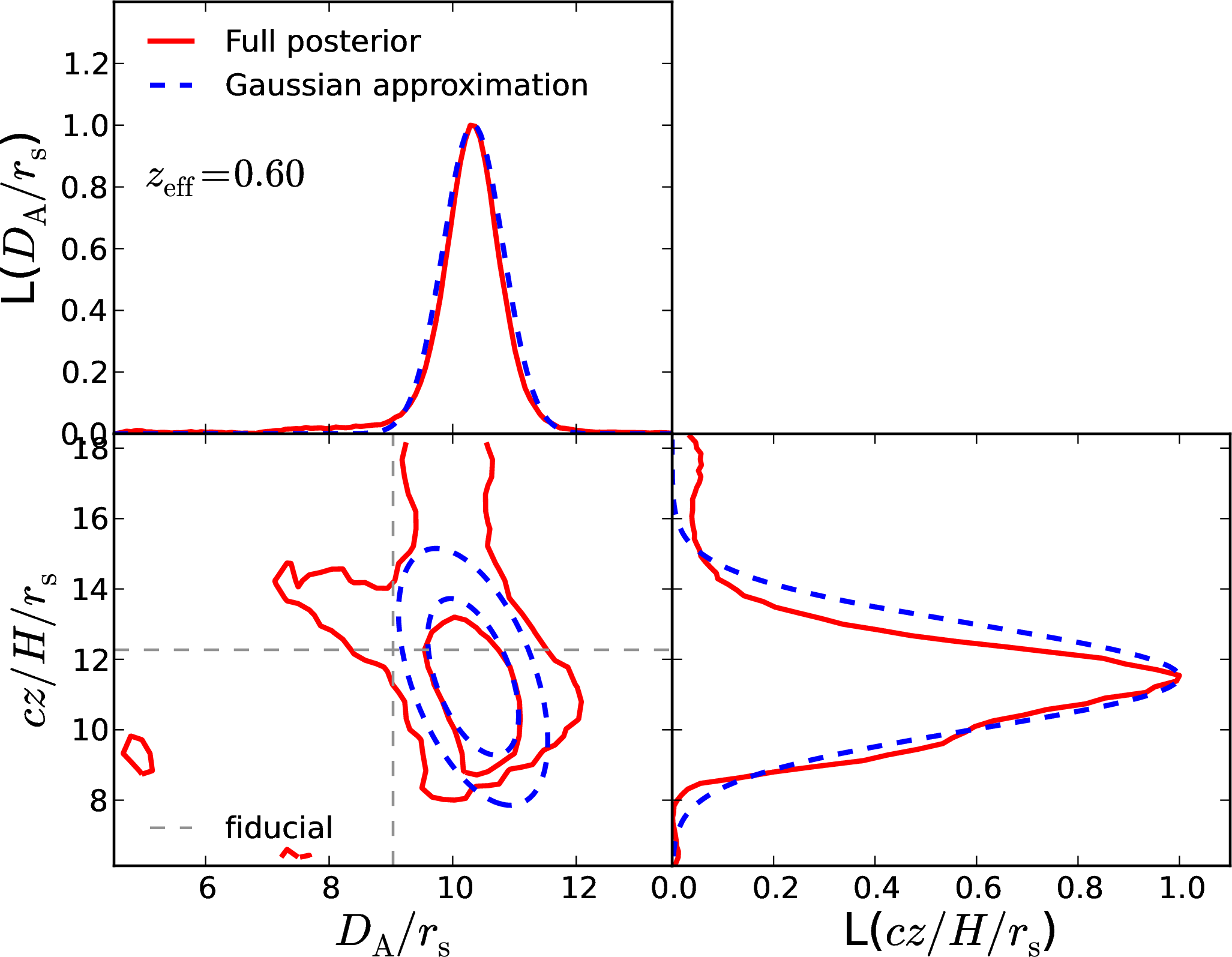}
\includegraphics[width=0.9\columnwidth]{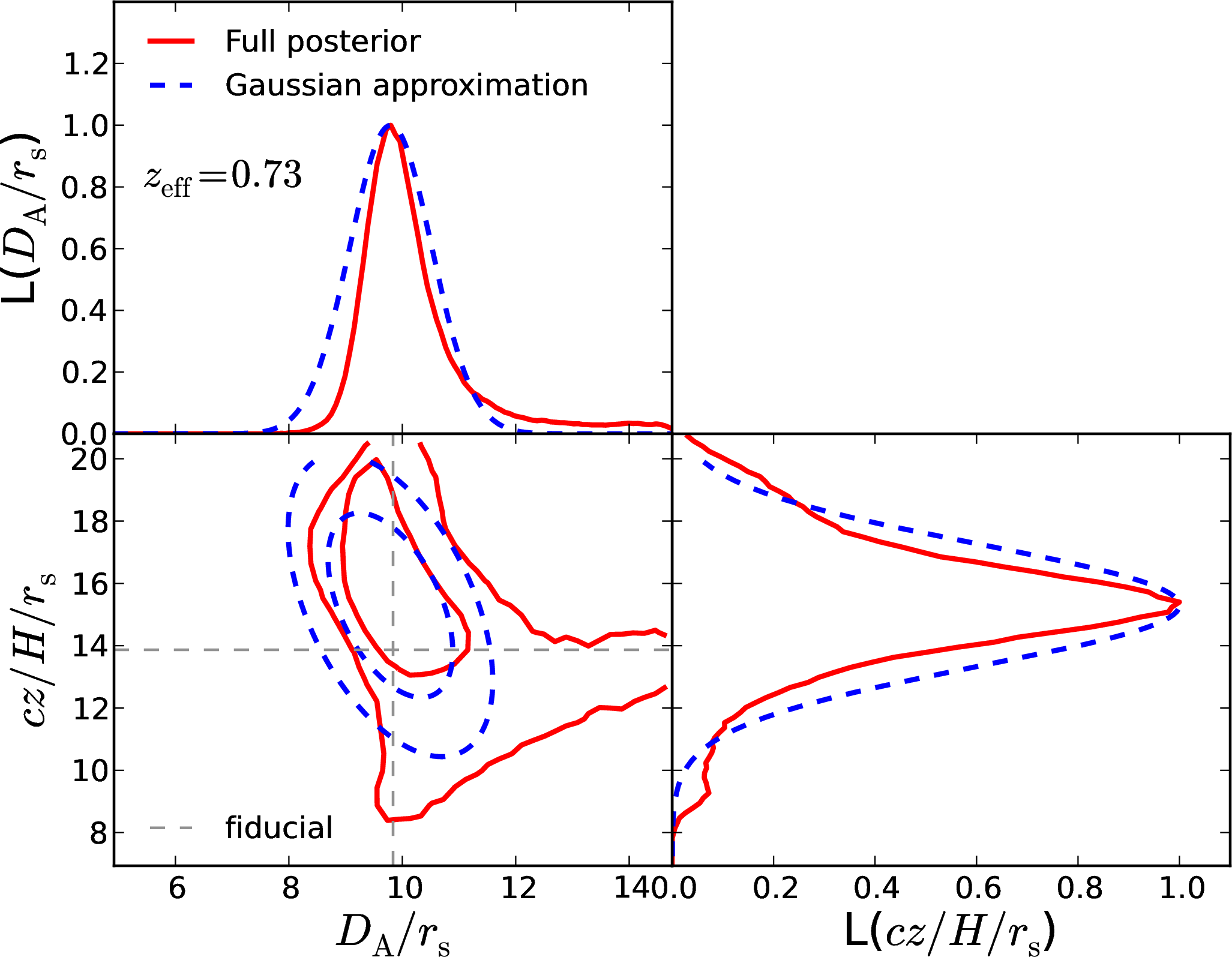}
\caption{ Marginalized posteriors of $cz/(Hr_{\rm s})$ and $D_{\rm A}/r_{\rm s}$ (red solid) obtained with WiggleZ post-reconstruction $\xi_{\perp, ||}$ in the three redshift bins $0.2<z<0.6$ (upper), $0.4<z<0.8$ (mid), $0.6<z<1.0$ (lower), using a flat prior on $\epsilon$ [-0.15,0.15].  The blue dashed lines are the Gaussian approximation when using the mode values, mean of the 68$\%$ CL regions and the cross-correlation $r$. The 2D even contours are the $68\%$ and $95\%$ CL regions and the thin gray dashed line marks the fiducial cosmology.  In the lowest redshift bin the data are not sufficient to constrain these parameters well.}
\label{fig:HDA_z26_epsilon0.15}
\end{center}
\end{figure}

Figure \ref{fig:HDA_z26_epsilon0.15} displays the posterior
probability distributions of $cz/(H \,r_{\rm s})$ and $D_{\rm A}/r_{\rm
  s}$. In the 2D panels the solid red contours indicate $68\%$ and
$95\%$ confidence level regions, and we indicate a Gaussian
approximation in each panel based on the statistics of the full
probability distributions.  It is apparent that the BAO-only analysis
of the $\Delta z^{\rm Far}$ and $\Delta z^{\rm Mid}$ samples yield
reasonable distance constraints, whereas the data in the $\Delta
z^{\rm Near}$ bin lacks the constraining power needed to draw
significant conclusions.  Table~\ref{tab:hda_wigglez_reconstructed} lists our resulting measurements of
$D_{\rm A}/r_{\rm s}$ and $cz/(H \, r_{\rm s})$.

To quantify the significance of detection of the anisotropic baryonic
feature in the WiggleZ clustering wedges we compared $\chi^2$ results
obtained with best-fit models using a $\Lambda$CDM-based template and
a ``no-wiggles'' template ($\Delta\chi^2 \equiv \chi^2_{\rm min,
  no-wiggle} - \chi^2_{\rm min, \Lambda CDM}$).  In this procedure,
for each model we vary $H r_{\rm s}$ and $D_{\rm A}/r_{\rm s}$ and
marginalize over all other shape parameters, as explained in detail in
\S 6.1 of \citet{KazinSanchezCuesta2013}.  We find that the significance of
detection, defined as $\sqrt{\Delta\chi^2}$ to be $1.6$, $2.7$, and
$2.9$ for $\Delta z^{\rm Near}$, $\Delta z^{\rm Mid}$, and $\Delta
z^{\rm Far}$, respectively.  Applying our pipeline to the WizCOLA
simulations, we find our results are consistent with the range of
expectations.

\begin{figure}
\begin{center}
\includegraphics[width=\columnwidth]{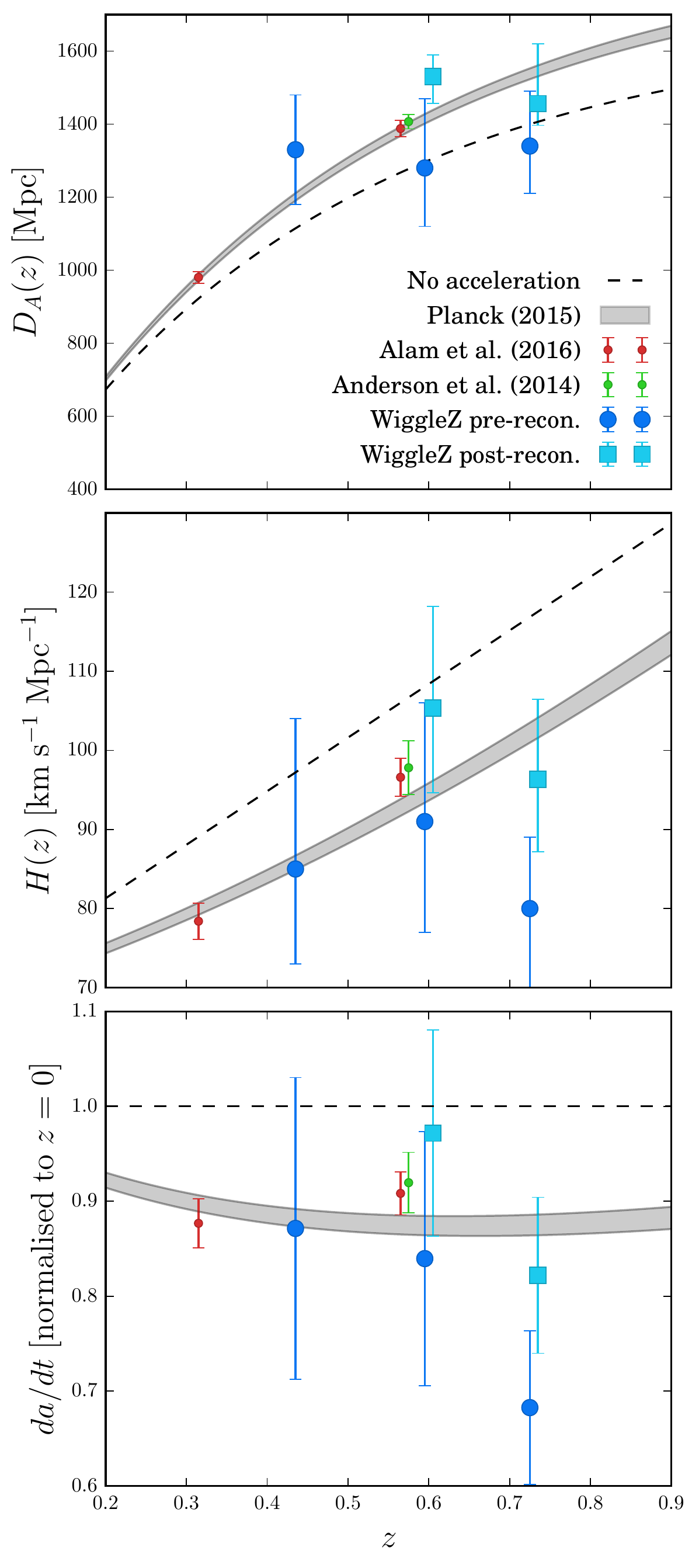}
\caption{The WiggleZ results of $D_{\rm A}(z)$, $H(z)$ and $da/dt$ for the pre-reconstruction results (dark blue circles) and post-reconstruction results (light blue squares). This is plotted against results from BOSS (both the results from \citet{AndersonAubourg2014} in green and \citet{AlamAta2016} in red), all with 68$\%$ CL regions in the $y$-axis (and redshift range on the $x$). The thick gray bands show the $1\sigma$ contour using Planck final constraints (TT,TE,EE+lowP+lensing+ext) \citep{Planck2015Parameters}. The dashed line indicates a cosmology with no acceleration (assuming Planck $H_0$).}
\label{fig:hubble_diagram}
\end{center}
\end{figure}

The results we find fitting the reconstructed wedges are consistent
with both prior WiggleZ studies, BOSS constraints from \citet{AndersonAubourg2014}, and
Planck cosmology \citep{Planck2015Parameters}, as illustrated in Figure \ref{fig:hubble_diagram}.

\section{Discussion and conclusion}

\label{sec:disc}
\label{sec:conclusion}

We have presented the first measurement of the 2D BAO signal in the WiggleZ Dark Energy Survey data \citep{KazinKoda2014}, where we fit for the cosmological parameters $\Omega_c h^2$, $D_A(z)$, and $H(z)$ for the three redshift bins $z \in \left[0.44, 0.60, 0.73\right]$.  Our final pre-reconstruction constraints appear in Table~\ref{tab:wigglezBinsParams}.  These results are consistent with the Flat $\Lambda$CDM cosmology derived from best-fitting Planck cosmological values and with previous large-scale structure measurements. Post-reconstruction results can be found in Table \ref{tab:hda_wigglez_reconstructed}, and are also consistent with best-fitting Planck cosmology. Pre- and post-reconstruction results are consistent, with there only being slight tension ($<2\sigma$) between results in the $z=0.60$ redshift bin. However, as the fitting methods make use of different data (full shape vs BAO peak), and the post-reconstruction likelihoods are highly non-Gaussian (as evidenced by Figure \ref{fig:HDA_z26_epsilon0.15}), the disagreement is smaller than the error bars might suggest, and is within the bounds of reasonable statistical fluctuation.

The constraints given by this analysis provide an important high-redshift consistency check against BOSS results as given in \cite{AndersonAubourg2014DR11}, who reported for their $z=0.57$ redshift bin, constraints of $D_A= (1421\pm20\, {\rm Mpc}) (r_d/r_{d,{\rm fid}})$ and $H = (96.8\pm3.4 \,{\rm km}\,{\rm s}^{-1}\,{\rm Mpc}^{-1})(r_d/r_{d,{\rm fid}})$. We find results consistent with the BOSS analysis, and show the BOSS constraints alongside our constraints and Planck cosmology in Figure~\ref{fig:hubble_diagram}. During the preparation of this manuscript, BOSS released a new analysis in \citet{AlamAta2016}, which are also consistent with the results we find. The larger uncertainty in our measurements compared to the BOSS is as-expected from the relative sizes of the data sets.  Nevertheless our results show that using a different type of galaxy tracer with much lower bias (bright blue galaxies as opposed to luminous red galaxies), we recover the same standard cosmological model. 

The main results of this analysis can be summarised as follows:
\begin{itemize}
	\item We update the unreconstructed 1D BAO measurement from \citet{BlakeKazin2011} using a more accurate covariance matrix based on WizCOLA mocks instead of lognormal realisations. The new best-fit parameters are consistent with the original measurements, with the maximum shift occurring in the highest redshift bin, whose value moved by slightly over 1$\sigma$ bringing it closer in line with the other two bins. See Table~\ref{tab:blakekazintable} for results.  Our results represent the final 1D BAO measurement using the {\em unreconstructed} WiggleZ data.  The most precise 1D BAO measurement from WiggleZ uses the {\em reconstructed} WiggleZ data as found in Kazin et al. (2014), which represents the final WiggleZ constraints from an angle-averaged BAO analysis.
	\item We validated our methodology by fitting 600 realisations of the WiggleZ survey generated by the WizCOLA simulations \citep{KodaBlake2015}. Our analysis recovered the input parameters of the simulation with no evidence for systematic bias. We also validate our methodology by testing agreement of cosmological parameters when analysing the 1D BAO signal with  \citet{BlakeKazin2011}.
	\item We thoroughly tested subtle methodological differences that could possibly have effected our analysis, such as different ways to combine the data from redshift bins, varying or fixing $\sigma_v$, or including the hexadecapole, which all gave consistent results.
	\item We performed the first cosmological analysis using the 2D BAO measurement of WiggleZ data using both pre and post reconstruction techniques. We detect the 2D BAO peak at a significance of slightly over 2$\sigma$ in each redshift bin for pre-reconstruction results, and approximately 3$\sigma$ for the $z=0.6$ and $z=0.73$ redshift bins for the reconstructed results, with the $z=0.44$ bin unable to provide convincing constraints. The best fit values of $\Omega_c h^2$, $H(z)$ and $D_A(z)$ for the pre-reconstruction fits are shown in Table~\ref{tab:wigglezBinsParams} and Fig.~\ref{fig:wigglezBinsMP}.  The results for $H(z)$ and $D_A(z)$ for post-reconstruction fitting are given in Table \ref{tab:hda_wigglez_reconstructed} and in Figures \ref{fig:HDA_z26_epsilon0.15} and \ref{fig:hubble_diagram}. These results are consistent with previous WiggleZ results, BOSS, and best fitting Planck cosmology.
\end{itemize}

\section*{Acknowledgments}

We gratefully acknowledge the input of the many researchers that were consulted during the creation of this paper. This work was supported by the Flagship Allocation Scheme of the NCI National Facility at the ANU. Parts of this research were conducted by the Australian Research Council Centre of Excellence for All-sky Astrophysics (CAASTRO), through project number CE110001020. This research has made use of NASA's Astrophysics Data System. We would also like to thank Joshua Calcino, Carolyn Wood and Sarah Thompson for their input. 

\clearpage
\bibliography{bibliography}

\appendix

	\section{Dewiggling Process} \label{app:dewiggle}
	
	In the literature the \verb;tffit; algorithm developed by \citet{EisensteinHu1998} is the most common method used to generate a power spectrum without the BAO feature. However, the use of this algorithm necessarily constrains an analysis to not only the precision of the algorithm, but also to the cosmologies considered when the algorithm was developed. Whilst most changes in cosmological models have been subtle in the past decade, a quick inspection of the changelog for CAMB\footnote{\url{http://camb.info/readme.html}} \citep{Lewis2000} shows over fifty software releases since the publication of the \verb;tffit; algorithm -- representing a continual divergence between CAMB and \verb;tffit; as CAMB continues to become more accurate and consistent with modern cosmological models, whilst \verb;tffit; remains static. 
	
	Given these reasons, we decided to develop an alternate method for generating a power spectrum without the BAO feature present. Given the regular updating of the CAMB software, a replacement algorithm would be most useful if it was capable of taking a standard linear power spectrum from CAMB and returning a filtered version, such that any changes in future cosmology would be reflected in the no wiggle power spectrum simply due to its presence in the original linear power spectrum from CAMB. To this end, several different methods of filtering power spectra were investigated, implemented, and tested, and we summarise those efforts here.  For more detail see \citet{HintonThesis2015}. 
	
	\subsection{Comparison of methods}
	
\begin{figure*}
		\begin{center}
			\includegraphics[width=0.7\textwidth]{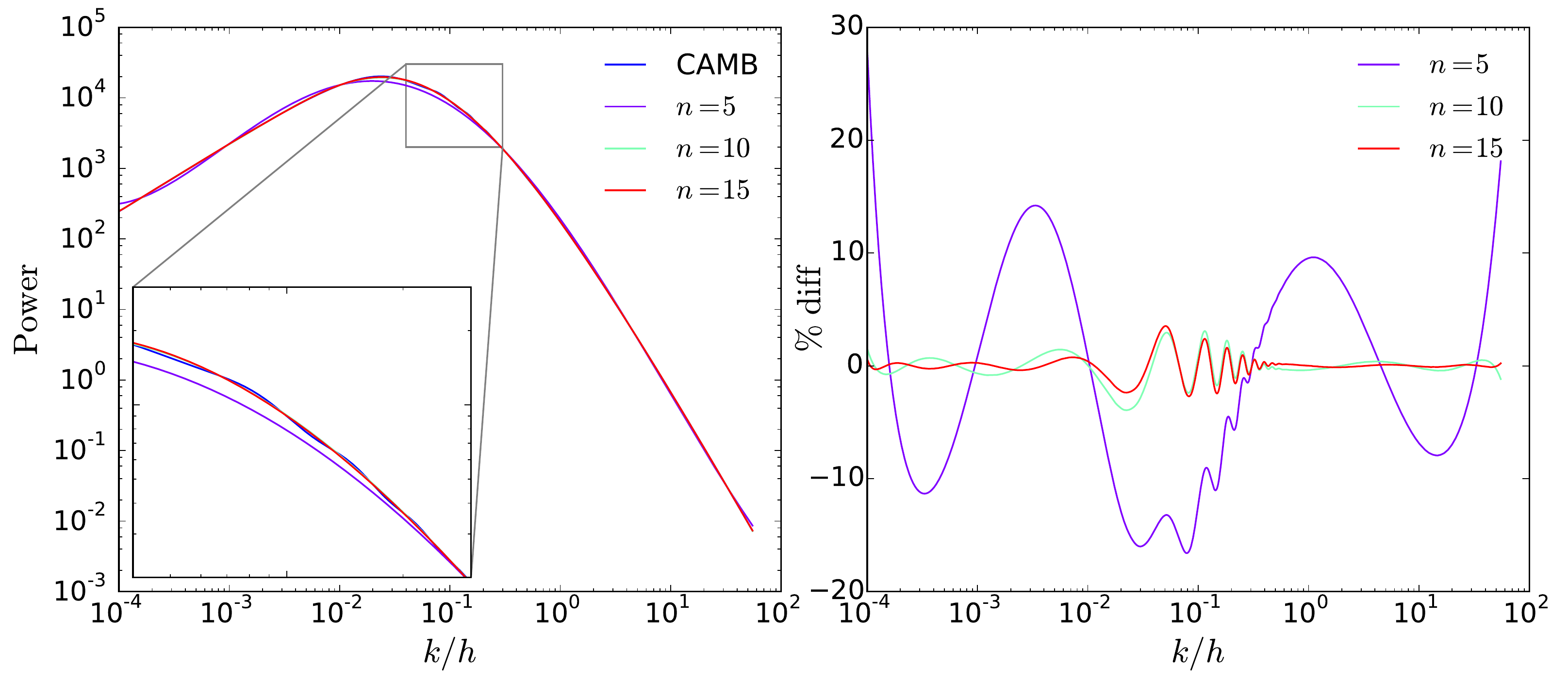}
			\caption{A comparison of the effects of increasing polynomial weight. Due to the high number of data points in the linear CAMB model ($>600$), even a high degree polynomial such as the 15 degree polynomial displayed in red, does not attempt to recover the BAO signal. Given the range of $k_*$ values typically used in model fitting, the right hand side of the graph where $k/h > 0.1$ is most relevant. It is desired that the polynomial fit converge to the CAMB power spectrum at high $k/h$, as occurs with higher order polynomial fits.}
			\label{fig:ApolyDegree}
		\end{center}
	\end{figure*}

	The BAO signal is present in the linear power spectrum generated by CAMB in the form of small scale oscillations after the main power peak, as illustrated in Figure~\ref{fig:ApolyDegree}. 
	
	Given the BAO signal is of small amplitude and restricted periodicity, both polynomial data fitting, low order spline interpolation, and frequency based filtering are all viable candidates for investigation.
	We found that low-pass and band-stop filters both failed because the strong broad range signal present in the power spectrum means that signal remains present at all frequencies, and thus there were no viable filters that extracted only the BAO signal.  The two methods that were successful were polynomial regression and spline interpolation.  
	
	\subsubsection{Polynomial regression}
	
	Polynomial regression is a tried and tested method for determining broad shape in a given spectrum \citep{baldry2014galaxy}. The higher order the polynomial fit becomes, the better the broad band shape extraction becomes, at the cost of eventually, as one keeps increasing the order, the polynomial model becomes detailed enough it begins to recover BAO signal. To counter this, one can introduce weights on the points, where the data points in the range of the BAO wiggle are down-weighted. To make this method more viable, a specific $k/h$ is not chosen as the centre point (as this strongly removes our model independence), instead we can note that the wiggle will appear approximately at the data peak, and down weight this area using a Gaussian weighting function, such that the weights supplied to the polynomial regression take the form $w = 1 - \alpha \exp\left(-k^2/2 \sigma^2\right)$. Using this, we can construct an array of polynomial fits where the polynomial degree, Gaussian width and amount of down-weighting are varied to determine the most effective construction to remove the BAO signal. In order to take advantage of the smooth shape of the power spectrum in the log domain, the polynomial regression is applied to the logarithm of the power spectrum.

	\begin{figure*}
		\begin{center}
			\includegraphics[width=0.7\textwidth]{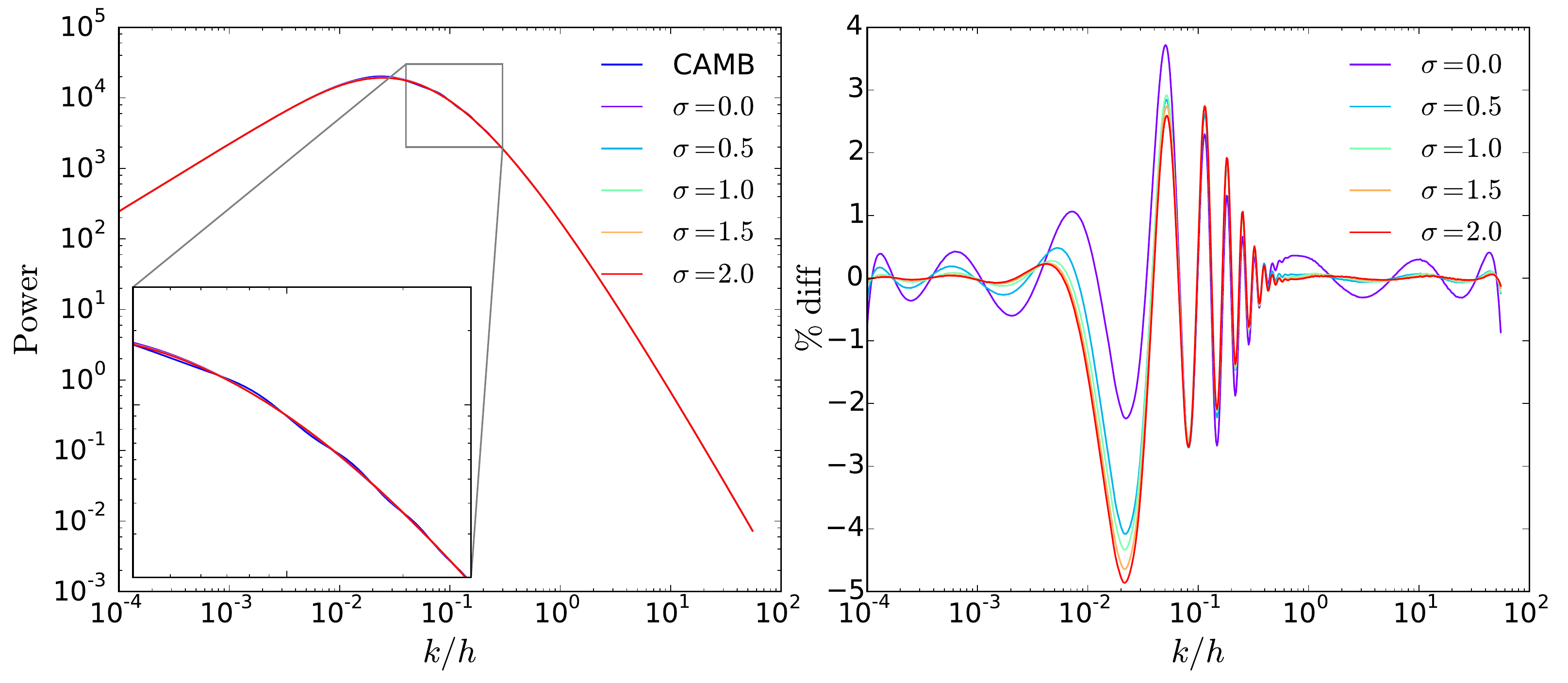}
			\caption{With polynomial degree fixed to $n = 13$, the width of the Gaussian used to down weight the peak of the spectrum is compared in this plot. It can be seen that no Gaussian ($\sigma= 0.0$) results in oscillations at high $k/h$, whilst the increasing $\sigma$ initially leads to better convergence at high $k/h$, with continually increasing $\sigma$ reducing the completeness of the BAO signal subtraction.}
			\label{fig:ApolySigma}
		\end{center}
	\end{figure*}
	\begin{figure*}
		\begin{center}
			\includegraphics[width=0.7\textwidth]{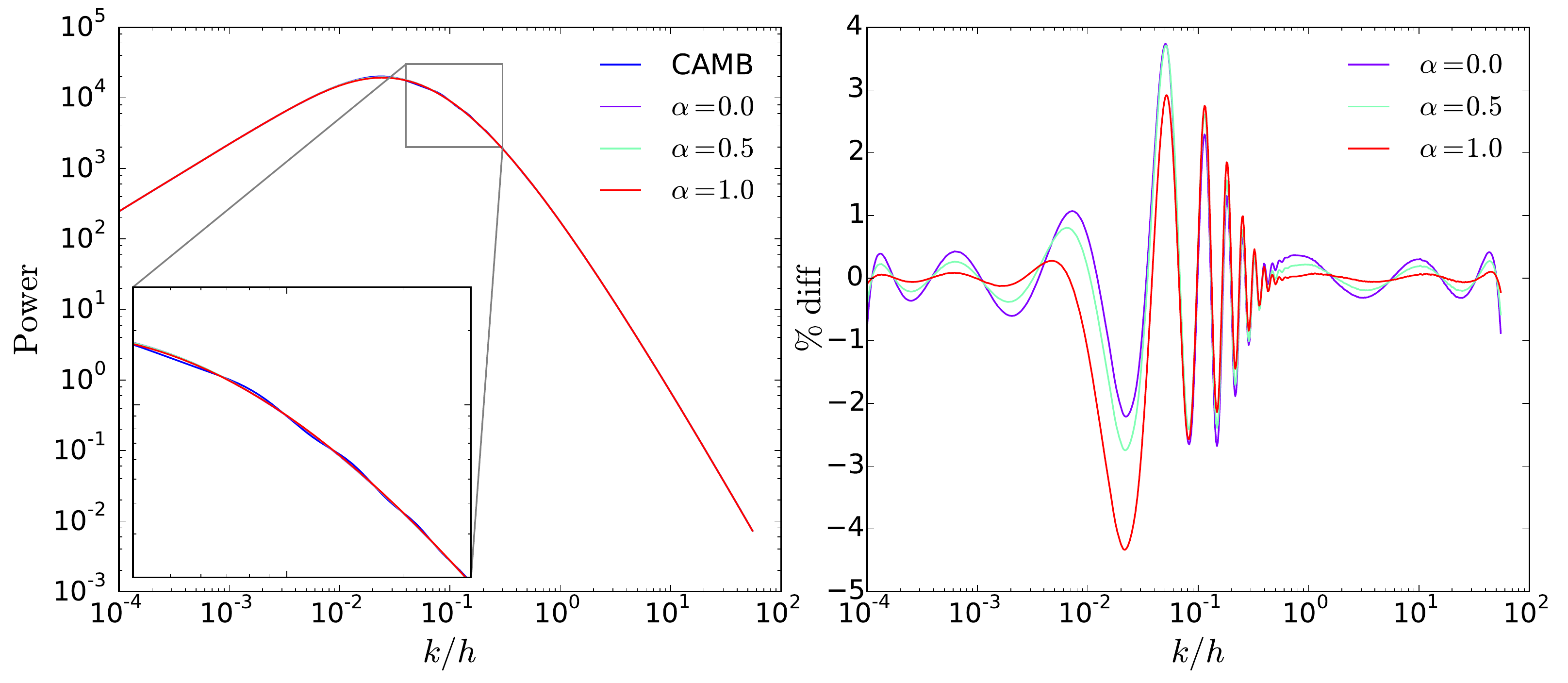}
			\caption{Setting $\sigma = 1$, we can examine the effect of the weight $\alpha$ of the Gaussian down weighting. As expected, setting the weight to zero gives the oscillations at high $k/h$ found in Figure \ref{fig:ApolySigma}. Setting the subtraction to full strength with $\alpha = 1.0$, we see that there is a downward shift in the polynomial fit (as the peak which lifts the fit has effectively been removed). Thus a compromising value in between must be chosen.}
			\label{fig:ApolyWeight}
		\end{center}
	\end{figure*}

	By comparing a wide array of parametrisations of polynomial degree $n$, Gaussian width $\sigma$, and Gaussian weight $\alpha$, a final combination of $n=13, \sigma=1, \alpha=0.5$ was chosen to act as the best choice for both strong BAO signal subtraction and non-distortion of the original linear power spectrum.

	\subsubsection{Spline Interpolation}

	The final method of removing the BAO signal from the linear power spectrum investigated was using spline interpolation. Similarly to the polynomial fits, it has the option of being supplied relevant weights for each data point, and thus a similar investigation as to weights was carried out for spline interpolation as was carried out for polynomial fitting. The spline fitting was found to be completely insensitive to modified weights, but highly sensitive to the positive smoothing factor $s$. A value of $s = 0.18$ compromises between BAO subtraction and low levels of distortion at high $k/h$, as determined by minimising the difference between the resultant spline model and the output of \verb;tffit;. Spline interpolation was similarly investigated in \citet{ReidPercival2010}, who found that use of a cubic b-spline with eight nodes fitted to $P_{\rm{lin}}(k) k^{1.5}$ produced likelihood surfaces in high agreement with formula from \citet{EisensteinHu1998}. In testing this methodology for potential use, no benefit was found to come from rotating the power spectrum via the $k^{1.5}$ in our algorithm. This was found for both tests using a univariate spline and a b-spline, however the similarity between the results of the different splines was such that only the univariate spline is documented.

	\subsection{Selection of final model}
	
\begin{figure}
		\begin{center}
			\includegraphics[width=\columnwidth]{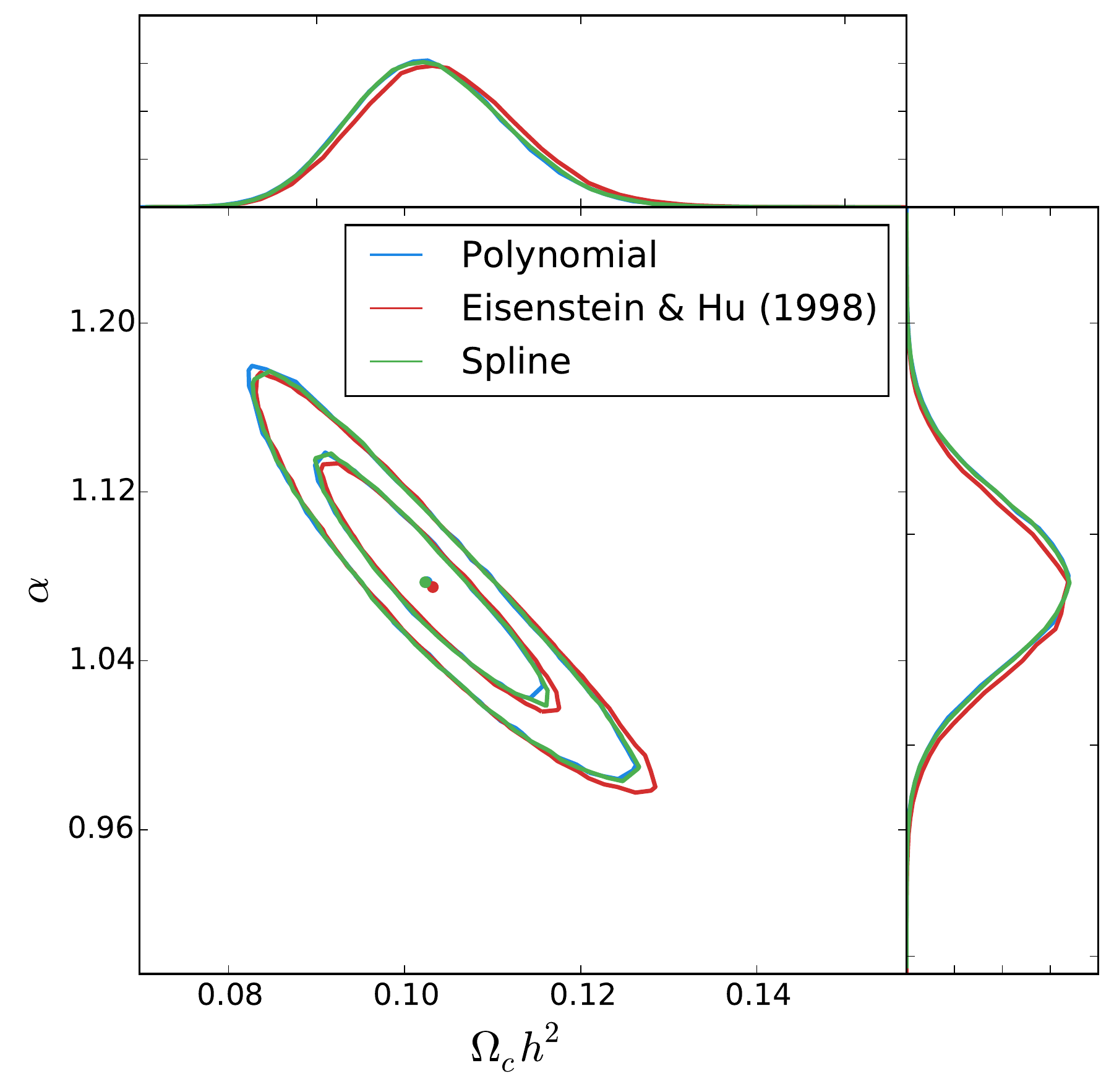}
		\end{center}
		\caption{A cosmological sensitivity test between the algorithm from \citet{EisensteinHu1998}, polynomial fitting and spline fitting. Likelihood surfaces and marginalised distributions were calculated using the WizCOLA simulation data at the $z=0.6$ redshift bin, where all 600 realisations have been used as input data, and $k_*$ fixed to $0.1$. With the low value of $k_*$ to increase the significance of the dewiggling algorithm and high data quality to reduce statistical uncertainty beyond the scope of the WiggleZ dataset, any deviation between the different methodologies should be represented by changes in the likelihood surface. However, as all likelihood surfaces agree to a high degree, we can conclude any difference in methodology is negligible in comparison to statistical uncertainty.}
		\label{fig:AcosmologyTest}
	\end{figure}

	Selecting the final method of dewiggling input spectra was done via looking explicitly at how the spectra are used in cosmological fitting: they are transformed into correlation functions and compared to observed data points. As such, the chosen optimal configurations for the polynomial and spline method were compared to \verb;tffit; by performing a cosmological sensitivity test wherein fits to WizCOLA data using the polynomial method, spline method and the algorithm given by \citet{EisensteinHu1998} are directly compared. To ensure this is robust, the value $k_*$ is fixed to 0.1, representing a fit with a very high level of dewiggling (hard thresholds are often limited to around this value, ie \citet{ChuangWang2012} have minimum $k_* = 0.09$), whilst still preserving some of the BAO peak with which to match. This analysis is given in Figure \ref{fig:AcosmologyTest}, and shows that for both spline and polynomial methods outlined above, statistical uncertainty in fits far exceeds any difference in matching results due to the change in dewiggling process. The polynomial method was selected to be the final method due to computational speed.

\section{Optimising range of scale included in fit} \label{app:truncation}

\begin{table}
	\centering
	\caption{A comparison of data fitting ranges found in prior literatures.}
	\label{tab:truncation}
	\begin{tabular}{lll}
		\specialrule{.1em}{.05em}{.05em} 
		Study & Range $(h^{-1}$Mpc) \\
		\specialrule{.1em}{.05em}{.05em} 
		\citet{XuPadmanabhan2012}    &     $30 < s < 200$           \\
		\citet{SanchezScoccola2012}  &    $40 < s < 200$             \\
		\citet{Sanchez2009}          &    $40 < s < 200$        \\  
		\citet{Gaztanaga2009}        &       $20 < s$    \\  
		\citet{ChuangWang2012}       &       $40 < s < 120$  \\  
		\citet{EisensteinZehavi2005} &       $10 < s < 180$          \\  
		\citet{BlakeDavis2011}       &       $10 < s < 180$         \\  
		\citet{KazinSanchezBlanton2012}&       $40 < s < 150$       \\  
		\citet{BlakeDavis2011}       &       $30 < s < 180$  \\  
		\citet{BlakeDavis2011}       &       $50 < s < 180$    \\  
		This work				& 	$25 < s < 180$ \\
		\specialrule{.1em}{.05em}{.05em} 
	\end{tabular}
\end{table}

\begin{figure}
	\begin{center}
		\includegraphics[width=\columnwidth]{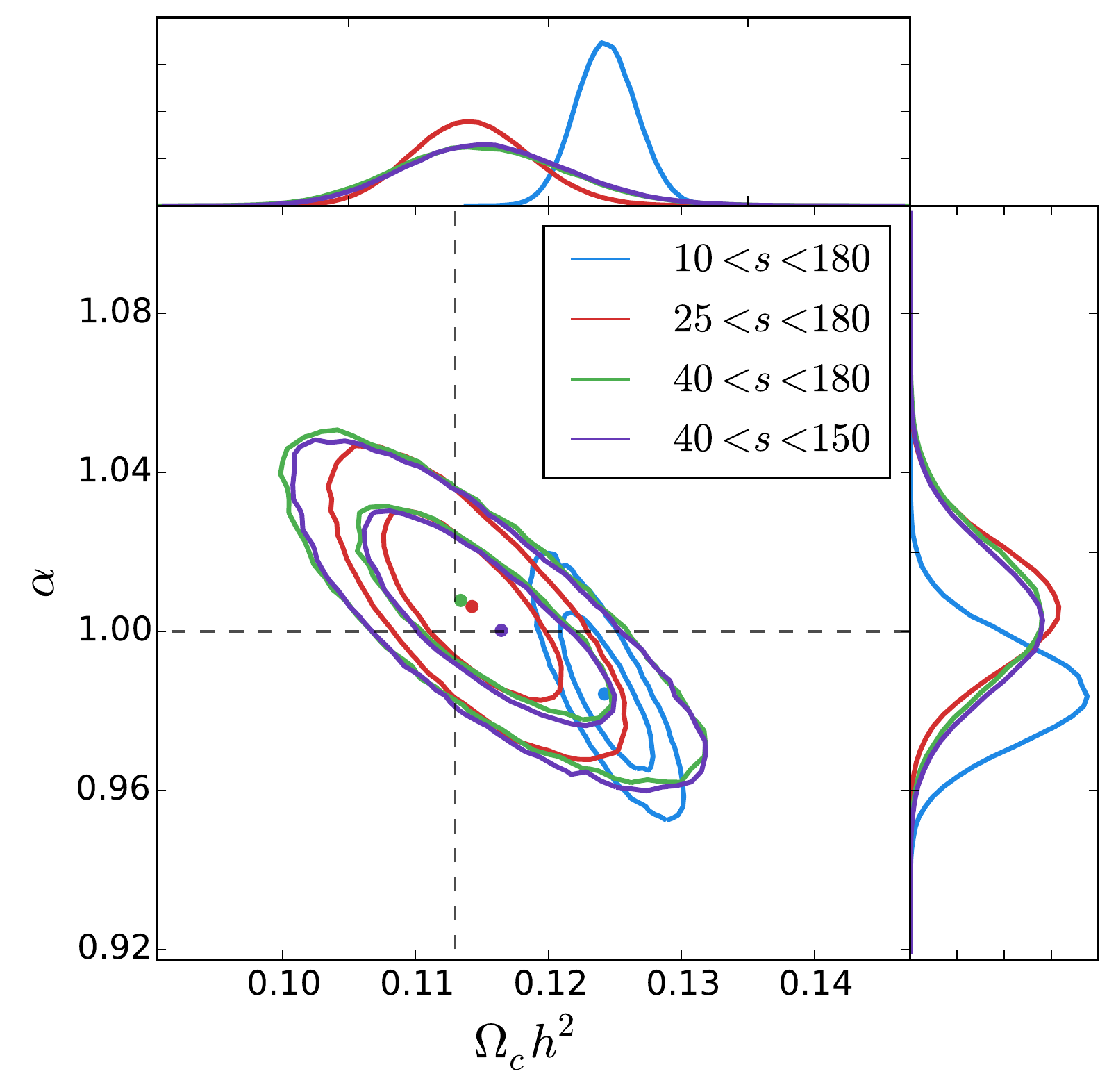}
	\end{center}
	\caption{We show four different dataset truncation values fit to the WizCOLA $z=0.6$ mean dataset. Utilising the $10<s<180 h^{-1}$ Mpc range employed by \citet{BlakeDavis2011} provided strong constraints on the parameters $\Omega_c h^2$ and $\alpha$, but recovered values more than $3-\sigma$ away from the desired outcomes (away from the known parameters used to create the simulation). Increasing the lower bound of the data shifted the recovered parameters to be well below $1-\sigma$ deviation from the desired outcome, at the cost of larger uncertainty in the likelihood surfaces. A reduced upper bound was tested as well due to its presence in prior literature, however minimal impact was found by reducing the upper limit. We increase the lower bound until we find unbiased parameter recovery at $s > 25\,h^{-1}\,$Mpc, and find the upper bound to be relatively insensitive to change, and fix it to $s < 180\,h^{-1}\,$Mpc. }
	\label{fig:CdatasetTrunc}
\end{figure}

The failure of correlation function models at small separations and their similarity at large separations mean it is important to evaluate the range of scales to include in the fit to the BAO signal, as detailed in \S\ref{sec:trunc}.  As the optimal data ranges vary depending on the survey volume, number density, and tracer bias, we investigate the effect of selecting different $s$-ranges on the recovered parameters when fitting to the WizCOLA simulation data. In order to constrain statistical uncertainty as much as possible, fits were performed to the combined dataset, in which the input values are determined from the mean of all 600 realisations of the WizCOLA simulation.  We then compare the output $\Omega_c h^2$ and $\alpha$ with the simulations as a function of the scales fitted. These are shown in Figure \ref{fig:CdatasetTrunc}, and the outcome of the comparison is the decision to use a dataset range of $25 < s < 180 \ h^{-1}$ Mpc.  We compare this range to previous analyses in Table \ref{tab:truncation}.

\section{Pre-reconstruction correlation} \label{app:cor}

Correlation coefficients corresponding to the outputs reported in Table~\ref{tab:wigglezBinsParams}, namely correlations between $\Omega_c h^2$, $D_A(z)$, and $H(z)$ across the three redshift bins, are given in Table~\ref{tab:cor}.  These were calibrated using WizCOLA mocks.  We chose to report these in terms of $D_A(z)$ and $H(z)$, we could equivalently have reported the correlations in $\alpha$ and $\epsilon$.  To convert between the two one would use equations \ref{eq:alpha1} and \ref{eq:alpha2}.

We note that the values reported in Table~\ref{tab:wigglezBinsParams} are determined using maximum likelihood statistics, and represent the best way to provide a summary statistic of the value as a data point. However, when combining these results by multivariate Gaussian approximation, the correct values to utilise change, and are given in Table~\ref{tab:gauss}.

\begin{table}
	\centering
	\caption{Gaussian approximation of parameter summaries in Table \ref{tab:wigglezBinsParams}. $D_A(z)$ given in units of Mpc, and $H(z)$ is presented in ${\rm km}\,{\rm s}^{-1} \, {\rm Mpc}^{-1}$.}
	\label{tab:gauss}
	\begin{tabular}{cccc}
		\hline
		Sample & $\Omega_c h^2$ &  $D_A$ &  $H$ \\
		\hline
		$0.2 < z < 0.6$ & $0.126\pm0.028$ & $1300\pm170$ & $90\pm16$ \\
		$0.4 < z < 0.8$ & $0.163\pm0.031$ & $1320\pm180$ & $95\pm16$ \\
		$0.6 < z < 1.0$ & $0.153\pm0.030$ & $1380\pm150$ & $81\pm10$ \\
		\hline
	\end{tabular}
\end{table}

\begin{table*}
	\centering
	\caption{Correlation values for the pre-reconstruction fits detailed in Table \ref{tab:wigglezBinsParams}. Redshift bins are placed in brackets after the parameter name.}
	\label{tab:cor}
	\begin{tabular}{rrrrrrrrrr}
		\hline
		~ & $\Omega_c h^2(0.44)$ &  $\Omega_c h^2(0.60)$ &   $\Omega_c h^2(0.73)$ &   $D_A(0.44)$ &   $D_A(0.60)$ &   $D_A(0.73)$ &    $H(0.44)$ &   $H(0.60)$ &   $H(0.73)$ \\
		\hline
	    $\Omega_m h^2(0.44)$ &      1    &        0.33 &        0.30 &       -0.19 &        0.02 &       -0.08 &        0.13 &        0.05 &        0.02 \\
	    $\Omega_m h^2(0.60)$ &      0.33 &        1    &        0.19 &       -0.07 &       -0.21 &       -0.03 &        0.06 &        0.20 &        0    \\
	    $\Omega_m h^2(0.73)$ &      0.30 &        0.19 &        1    &       -0.05 &       -0.01 &       -0.28 &       -0.05 &        0.08 &        0.20 \\
	    $D_A(0.44)$          &     -0.19 &       -0.07 &       -0.05 &        1    &        0.01 &        0.04 &        0.12 &       -0.01 &        0.03 \\
	    $D_A(0.60)$          &      0.02 &       -0.21 &       -0.01 &        0.01 &        1    &       -0.01 &        0    &        0.10 &       -0.01 \\
	    $D_A(0.73)$          &     -0.08 &       -0.03 &       -0.28 &        0.04 &       -0.01 &        1    &        0.01 &        0    &        0.07 \\
        $H(0.44)$            &      0.13 &        0.06 &       -0.05 &        0.12 &        0    &        0.01 &        1    &        0.06 &        0.03 \\
	    $H(0.60)$            &      0.05 &        0.20 &        0.08 &       -0.01 &        0.10 &        0    &        0.06 &        1    &        0    \\
	    $H(0.73)$            &      0.02 &        0    &        0.20 &        0.03 &       -0.01 &        0.07 &        0.03 &        0    &        1    \\
		\hline
	\end{tabular}
\end{table*}

\bsp	
\label{lastpage}
\end{document}